\documentclass[10pt,preprintnumbers, floatfix, letterpaper, onecolumn,aps,prd,epsfig,nofootinbib,nat_bib,longbibliography]{revtex4-1}

\usepackage{graphicx}
\usepackage[utf8]{inputenc}
\usepackage[T1]{fontenc}
\usepackage{physics}
\usepackage{latexsym}
\usepackage{epstopdf}
\usepackage{latexsym}
\usepackage{amssymb}
\usepackage{amsmath}
\usepackage{mathtools}
\usepackage{color}
\usepackage{mathrsfs}
\usepackage{xparse}
\usepackage{enumerate}
\usepackage{multirow,tabularx,boldline,array}
\usepackage{bbding}
\usepackage{enumitem}
\usepackage{pifont}
\usepackage[center]{subfigure}
\usepackage[
            pdfstartview=FitH,
            bookmarksnumbered=true,
            bookmarksopen=true,
            colorlinks,
            linkcolor=blue,
            anchorcolor=green,
            citecolor=blue
            ]{hyperref}
\begin{document}

  \renewcommand\arraystretch{2}
 \newcommand{\bq}{\begin{equation}}
 \newcommand{\eq}{\end{equation}}
 \newcommand{\bqn}{\begin{eqnarray}}
 \newcommand{\eqn}{\end{eqnarray}}
 \newcommand{\nb}{\nonumber}
 \newcommand{\lb}{\label}
 \newcommand{\cb}{\color{blue}}
    \newcommand{\cc}{\color{cyan}}
        \newcommand{\cm}{\color{magent_a}}
\newcommand{\rc}{\rho^{\scriptscriptstyle{\mathrm{I}}}_c}
\newcommand{\rd}{\rho^{\scriptscriptstyle{\mathrm{II}}}_c}
\NewDocumentCommand{\evalat}{sO{\big}mm}{%
  \IfBooleanTF{#1}
   {\mleft. #3 \mright|_{#4}}
   {#3#2|_{#4}}%
}
\newcommand{\PRL}{Phys. Rev. Lett.}
\newcommand{\PL}{Phys. Lett.}
\newcommand{\PR}{Phys. Rev.}
\newcommand{\CQG}{Class. Quantum Grav.}


\title{Island and Page curve for one-sided asymptotically flat black hole}
\author{Wen-Cong Gan}
	\email{Wen-cong$\_$Gan1@baylor.edu}
	\affiliation{GCAP-CASPER, Physics Department,
Baylor University, Waco, Texas 76798-7316, USA}
	\affiliation{Department of Physics, Nanchang University, Nanchang, 330031, China}
	\author{Dong-Hui Du}
	\email{donghuiduchn@gmail.com,}
	\affiliation{School of Physics and Astronomy, Sun Yat-sen University, Guangzhou 510275, China}
	\affiliation{Department of Physics, Nanchang University, Nanchang, 330031, China}
		\author{Fu-Wen Shu}
	\email{shufuwen@ncu.edu.cn; Corresponding author}
	\affiliation{Department of Physics, Nanchang University, Nanchang, 330031, China}
	\affiliation{Center for Relativistic Astrophysics and High Energy Physics, Nanchang University, Nanchang,
330031, China}
\affiliation{GCAP-CASPER, Physics Department,
Baylor University, Waco, Texas 76798-7316, USA}
\affiliation{Center for Gravitation and Cosmology, Yangzhou University, Yangzhou, China}

\date{\today}

\begin{abstract}

Great breakthrough in solving black hole information paradox took place when semiclassical island rule for entanglement entropy of Hawking radiation was proposed in recent years. Up to now, most papers which discussed island rule of asymptotic flat black hole with $D \ge 4$ focus on eternal black hole. In this paper, we take one more step further by discussing island of ``in'' vacuum state which describes one-sided asymptotically flat black hole formed by gravitational collapse in $D \ge 4$. We find that island $I$ emerges at late time and saves entropy bound. And boundary of island $\partial I$ depends on the position of cutoff surface. When cutoff surface is far from horizon, $\partial I$ is inside and near horizon. When cutoff surface is set to be near horizon, $\partial I$ is outside and near horizon. This is different from the case of eternal black hole in which $\partial I$ is always outside horizon no matter cutoff surface is far from or near horizon. We will see that different states will manifestly affect $S_{\text{ent}}$ in island formula when cutoff surface is far from horizon and thus have different result for Page time. 

\end{abstract}

\maketitle
\tableofcontents
\section{Introduction}
 \renewcommand{\theequation}{1.\arabic{equation}}\setcounter{equation}{0}

Black hole information paradox is a long lasting debate over more than 40 years since Hawking discovered that information may be lost in evaporation of black hole \cite{Hawking:1976ra}. Hawking's calculation implies that von Newman entropy of Hawking radiation will increase monotonically. On the other hand, quantum mechanics requires black hole evaporation to be unitary, thus von Newman entropy of Hawking radiation should obey Page curve if information is preserved during black hole evaporation \cite{Page:1993wv}. Since then, producing Page curve in gravitational calculation is a key step towards solving information paradox.

\begin{figure}[h!]
    \begin{tabular}{cc}
\includegraphics[height=7.5cm]{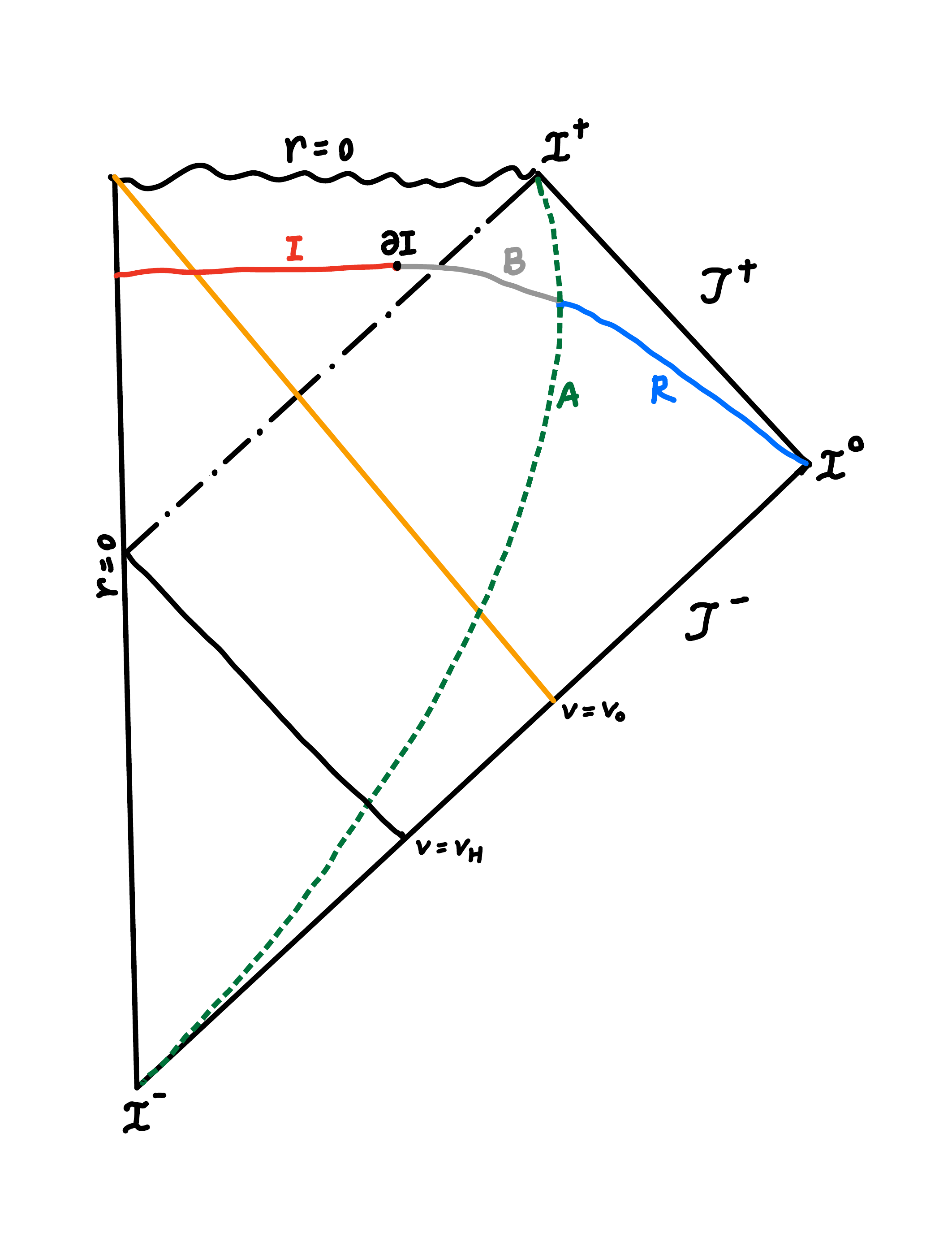}&
\includegraphics[height=7.5cm]{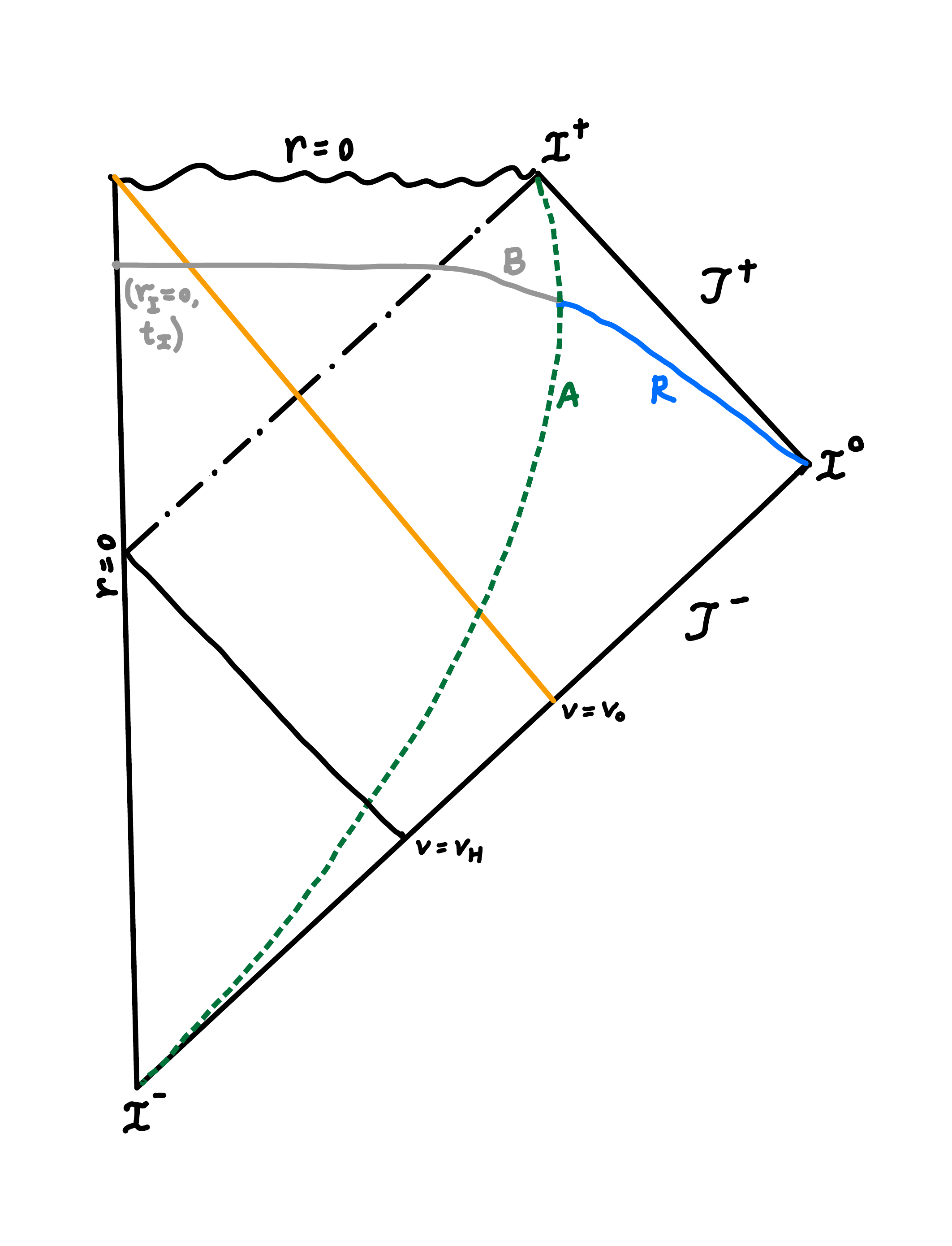}\\
	(a) & (b)  \\[6pt]
	\end{tabular}
\caption{(a) The Penrose diagram for one-sided black hole formed by spherical null shell collapsing at $v=v_0$ (orange line) \cite{Fabbri:2005mw}. Observer on cutoff surface $A$ (green dashed line) collects Hawking radiation in region $R$. Island (red line) penetrates into interior of black hole. $\partial I$ is quantum extremal surface. $I \cup B \cup R$ is a Cauchy slice. (b) Without island, we can equivalently fix $r_I=0$ and $t_I$ as some constant.}
\label{black hole}
\end{figure}

In \cite{Penington:2019npb,Almheiri:2019psf,Almheiri:2019hni}, {\it island rule} is proposed to calculate Page curve of Hawking radiation (see e.g. \cite{Almheiri:2020cfm} for a review). Island rule states that fine grained entropy of black hole is given by
\bqn\lb{bh}
S(BH)=\text{min}\left\{\text{ext}\left[\frac{\text{Area}(\partial I)}{4G_N}+S_{\text{ent}}(B)\right]\right\},
\eqn
and fine grained entropy of Hawking radiation is given by
\bqn\lb{r}
S(R)=\text{min}\left\{\text{ext}\left[\frac{\text{Area}(\partial I)}{4G_N}+S_{\text{ent}}(I\cup R)\right]\right\},
\eqn
where $R$ denotes the region outside cutoff surface $A$ and collecting Hawking radiation, $I$ is called island and denotes an codimension-one hypersurface which penetrates into the interior of black hole, $\partial I$ is the codimension-two boundary of $I$ (see fig.(\ref{black hole})).  $\partial I$ is chosen to be the quantum extremal surface (QES) \cite{Engelhardt:2014gca} that extremizes the generalized entropy \cite{Faulkner:2013ana,Lewkowycz:2013nqa,Dong:2016hjy}
\bqn
S_{\text{gen}}=\frac{\text{Area}(\partial I)}{4G_N}+S_{\text{ent}},
\eqn
where the first term is area term from Ryu-Takayanagi formula \cite{Ryu:2006bv,Hubeny:2007xt}, and the second term is coarse-grained entropy of matter fields. If there are more than one extremal surface, the global minimum one should be taken. This is the meaning of ``min'' and ``ext'' in eq.(\ref{bh}) and eq.(\ref{r}). Island formula can be derived by using replica trick to construct replica wormhole \cite{Penington:2019kki,Almheiri:2019qdq}.

Island formula was discussed in \cite{Almheiri:2019psf,Almheiri:2019hni,Chen:2019uhq} for $2D$ Jackiw-Teitelboim (JT) gravity in asymptotically anti-de Sitter (AdS) spacetimes coupled to a thermal bath, in \cite{Balasubramanian:2020xqf,Geng:2021wcq} for de Sitter (dS), in \cite{Hu:2022ymx} for curvature-squared gravity, in \cite{Ling:2020laa} for charged black holes, and in \cite{Almheiri:2019psy} for higher dimensions. And the case for asymptotically flat spacetimes was discussed in \cite{Gautason:2020tmk,Anegawa:2020ezn,Hartman:2020swn,Yu:2021cgi} for dilaton gravity, in \cite{Azarnia:2021uch} for flat-space cosmology, in \cite{Wang:2021mqq} for a family of $2D$ exactly solvable black holes, in \cite{Alishahiha:2020qza,Hashimoto:2020cas,Matsuo:2020ypv,Arefeva:2021kfx,Dong:2020uxp} for Schwarzschild black hole, in \cite{Kim:2021gzd,Wang:2021woy} for Reissner-Nordstr\"om black hole and in \cite{He:2021mst} for a family of eternal black holes. Due to rapid progress in this field, above is only an incomplete list of references.

Up to now, as far as we know, most of the papers that discuss Schwarzschild black hole \cite{Alishahiha:2020qza,Hashimoto:2020cas,Matsuo:2020ypv,Arefeva:2021kfx,Dong:2020uxp} and Reissner-Nordstr\"om black hole \cite{Kim:2021gzd,Wang:2021woy} only concern about eternal black hole and static vacuum \cite{Matsuo:2021mmi}. In \cite{Alishahiha:2020qza}, the authors discuss one-sided dynamical black hole but they implicitly assume Hartle-Hawking state \cite{Hartle:1976tp} which is more appropriate to eternal black hole. Hartle-Hawking state or Hartle-Hawking vacuum is the unique state which is regular in the whole Kruskal extension of Schwarzschild black hole and is invariant under Schwarzschild time translation. It represents a black hole in thermal equilibrium with an outgoing Hawking radiation reaching future null infinity $\mathcal{J}^+$ and an equal incoming thermal radiation coming in from past null infinity $\mathcal{J}^-$.

In this paper, we will discuss ``in'' vacuum state which describes one-sided dynamic asymptotically flat black hole formed from collapsing of spherical null shell (see fig.(\ref{black hole})) \cite{Fabbri:2005mw}. It can be approximated by Unruh state in late time limit. ``In'' vacuum state and Unruh state both can describe black hole formed by gravitational collapse since they both contain no incoming flux from $\mathcal{J}^-$. Hence, it is generally believed that, compared with the Hartle-Hawking state, ``in'' vacuum state is more appropriate to one-sided dynamic black hole. We will see that different state will manifestly affect $S_{\text{ent}}$ in island formula and thus have different result for Page time.  In addition, different from Hartle-Hawking state for eternal black hole,\footnote{ As shown in Appendix \ref{app-1}, $s$-wave approximation for eternal black hole in Hartle-Hawking state is questionable. Since now there are ingoing modes in addition to the outgoing modes. We would like to thank the anonymous referee for pointing out this issue.} due to absence of incoming flux, $s$-wave approximation is valid for ``in'' vacuum state of one-sided black hole formed from collapsing of spherical null shell when cutoff surface $A$ is far from horizon (see Appendix \ref{app-1}). The use of ``in'' vacuum state is the key difference between our paper and the other papers that use Hartle-Hawking state \cite{Alishahiha:2020qza,Hashimoto:2020cas,Matsuo:2020ypv,Arefeva:2021kfx,Kim:2021gzd,Wang:2021woy,He:2021mst,Matsuo:2021mmi}.

This paper is organized as follows. In Sec. \ref{sec-entropy}, entanglement entropy $S_{\text{ent}}$ is discussed and we will see its dependence on state. In Sec. \ref{4d-far}, we discuss the case in which cutoff surface is far from horizon and find that boundary of island $\partial I$ is inside black hole horizon and produce the Page curve. In Sec. \ref{4d-near}, we discuss the case in which cutoff surface is near horizon and find that boundary of island $\partial I$ is outside black hole horizon and produce the Page curve. In Sec. \ref{higher-d}, we discuss the case for higher dimensional ($D>4$) black hole formed by collapse of spherical null shell and find similar results as the $4D$ case. The conclusion and discussion are in Sec. \ref{discussion}.

\section{Entanglement entropy in black hole background}\lb{sec-entropy}
 \renewcommand{\theequation}{2.\arabic{equation}}\setcounter{equation}{0}

In this paper, we will mainly focus on the island of ``in'' vacuum state \cite{Fabbri:2005mw} in the background of Schwarzschild black hole. $I \cup B \cup R$ is a Cauchy slice on which the quantum state is pure since we consider vacuum state which is pure. Then
\bqn
S_{\text{ent}}(B)=S_{\text{ent}}(I\cup R),
\eqn
and thus $S(BH)=S(R)$ as shown in eqs. \eqref{bh} and \eqref{r}. In the following, we will mainly focus on $S_{\text{ent}}(B)$ for simplicity.

\subsection{Cutoff surface far from horizon}
 When the cutoff surface $A$ is far from horizon, it is reasonable to assume that we only consider $2D$ massless scalar fields,  i.e. $s$-wave approximation is valid (see Appendix \ref{app-1}).  Then the matter part $S_{\text{ent}}$ is approximately given by entanglement entropy of massless scalar fields in $2D$ spacetime \cite{Gautason:2020tmk}\footnote{ In Eq.(\ref{sent}), we omit UV cutoff parameter, since it can be absorbed in the renormalization of Newton constant $G_N$ \cite{Gautason:2020tmk,Almheiri:2019psf,Alishahiha:2020qza,Hashimoto:2020cas,Susskind:1994sm}.}
\begin{eqnarray}\lb{sent}
S_{\text{ent}}&=&\frac{c}{6}  \ln \left( d(A,I)^2 e^{\rho_A} e^{\rho_I}
\right)_{t_{\pm}=0}\\
&=&\frac{c}{12}  \ln \left( d(A,I)^4 e^{2\rho_A} e^{2\rho_I}
\right)_{t_{\pm}=0},
\end{eqnarray}
where $c$ is central charge. And
\begin{eqnarray}
d(A,I)=\sqrt{[x^+(A)-x^+(I)][x^-(I)-x^-(A)]}
\end{eqnarray}
is the distance between $A$ and $\partial I,$\footnote{Throughout the paper, we denote the coordinate of $\partial I$ by using subscript $I$ such as $(u_I,v_I)$.} in flat metric $ds^2=-dx^+ dx^-$, and
\begin{eqnarray}
(e^{2\rho})_{t_{\pm}=0}=e^{2\rho(x^+,x^-)}.
\end{eqnarray}

Eq.(\ref{sent}) can be understood as follows. Entanglement entropy in vacuum state of conformal field theory (CFT) \cite{Calabrese:2004eu,Calabrese:2009qy,Casini:2009sr} in flat spacetime $ds^2=-dx^+ dx^-$ is given by
\begin{eqnarray}\lb{sent-flat}
S_{\text{ent}}&=&\frac{c}{3}  \ln \left( d(A,I)
\right).
\end{eqnarray}
 Fields in Minkowski spacetime can be quantized as
\bqn
f=\int \frac{d \omega}{\sqrt{4 \pi \omega}} \left( a_{\omega} e^{-i\omega x^+}+a_{\omega}^{\dagger} e^{i\omega x^+}+b_{\omega} e^{-i\omega x^-}+b_{\omega}^{\dagger} e^{i\omega x^-} \right).
\eqn
Minkowski vacuum state is defined with respect to coordinates $(x^+,x^-)$ (i.e. with respect to the modes $(4 \pi \omega)^{-1/2} e^{-i\omega x^+}$ and $(4 \pi \omega)^{-1/2} e^{-i\omega x^-}$) such that
\bqn
 a_{\omega} |0\rangle=0, \; b_{\omega} |0\rangle=0.
\eqn
That is to say, vacuum expectation value (VEV) of normal ordered stress tensor in coordinates $(x^+,x^-)$ vanishes $\langle 0|:T_{x^+ x^+}:|0\rangle=\langle 0|:T_{x^- x^-}:|0\rangle=0$. 

In two-dimensional gravity $ds^2=-e^{2\rho(x^+,x^-)}dx^+ dx^-$, VEV of normal ordered stress tensor defines a set of function $t_{\pm}$,\footnote{Throughout the paper, we assume $c G_N^{(D)} \ll r_h^{D-2}$ for $D$-dimensional macroscopic black hole so that we can neglect backreaction effect for simplicity.}
\bqn
\langle \Psi|:T_{\pm \pm}(x^{\pm}):|\Psi\rangle \equiv -\frac{\hbar}{12\pi} t_{\pm}(x^{\pm}),
\eqn
where $T_{++}\equiv T_{x^+ x^+}$, and $T_{--}\equiv T_{x^- x^-}$. Under transformation $x^{\pm} \rightarrow y^{\pm}$,
\bqn
\rho(x^{\pm}) \rightarrow \rho(y^{\pm})= \rho(x^{\pm})+\frac{1}{2}\ln \left( \frac{dx^+}{dy^+}\frac{dx^-}{dy^-}\right),
\eqn
and
\bqn
t_{\pm}(y^{\pm})= \left(\frac{dx^{\pm}}{dy^{\pm}}\right)^2 t_{\pm}(x^{\pm})+\frac{1}{2}\{x^{\pm},y^{\pm} \},
\eqn
where $\{x^{\pm},y^{\pm} \}\equiv \left(\frac{d^3x^{\pm}}{dy^{\pm 3}}\right)/\left(\frac{dx^{\pm}}{dy^{\pm}}\right) - \frac{3}{2} \left(\frac{d^2 x^{\pm}}{dy^{\pm 2}}/\frac{dx^{\pm}}{dy^{\pm}}\right)^2$ is the Schwarzian derivative. Obviously, $t_{\pm}$ is a state-dependent and coordinate-dependent function.

After Weyl transformed from $ds^2=-dx^+ dx^-$ to $ds^2=-e^{2\rho(x^+,x^-)}dx^+ dx^-$, vacuum state $|0\rangle$ in flat spacetime is mapped to vacuum state $|vac\rangle$ which is still defined with respect to coordinates $(x^+,x^-)$ in curved spacetime. Then VEV of normal ordered stress tensor in coordinates $(x^+,x^-)$ also vanishes $\langle vac|:T_{x^+ x^+}:|vac\rangle=\langle vac|:T_{x^- x^-}:|vac\rangle=0$. Then $t_{\pm}=0$ for vacuum state $|\Psi\rangle=|vac\rangle$ in $(x^+,x^-)$ coordinates, thus $(e^{2\rho})_{t_{\pm}=0}=e^{2\rho(x^+,x^-)}$. In addition, under Weyl transformation, entanglement entropy is transformed as \cite{Almheiri:2019psf}
\bqn
S_{\Omega^{-2} g}=S_{g}-\frac{c}{6} \sum_{endpoints} \ln(\Omega).
\eqn
Setting $\Omega^{-2}=e^{2\rho}$, we have
\bqn
S_{e^{2\rho} ds^2}=S_{ds^2}+\frac{c}{6} \sum_{endpoints} \ln(e^{\rho}).
\eqn
Finally, eq.(\ref{sent-flat}) is transformed to eq.(\ref{sent}).

\subsection{Cutoff surface near horizon}
When length scale of the region is sufficiently small compared with the length scale of the curvature, entropy of matter fields can be approximately given by the one in vacuum state in flat spacetime no matter which spacetime and state we consider. Thus renormalized entanglement entropy of matter term is now given by \cite{Hashimoto:2020cas,Casini:2005zv}
\begin{eqnarray}\lb{sent2}
S_{\text{ent}}&=&-\kappa c \frac{\text{Area}}{L^2},
\end{eqnarray}
where $\kappa$ is a constant and $L$ is the geodesic distance between boundary of island $\partial I$ and cutoff surface $A$.  ``Area'' in \eqref{sent2} is given by the area of cutoff surface which is approximately equal to the area of $\partial I$.

 Since we assume both $\partial I$ and cutoff surface is near horizon, $L$ can be approximately given by \cite{Hashimoto:2020cas}
\bqn\lb{L}
L &=& \int \sqrt{-e^{2\rho(x^+,x^-)}dx^+ dx^-} \\
&\approx& \sqrt{[x^+(A)-x^+(I)][x^-(I)-x^-(A)]e^{\rho_A} e^{\rho_I}}\\
&=&\sqrt{d(A,I)^2 e^{\rho_A} e^{\rho_I}}.
\eqn
We would like to point out that eq.(\ref{L}) is approximately valid when $L$ is sufficiently small with respect to the length scale of the curvature.

\section{Cutoff surface far from horizon}\lb{4d-far}

Now let us consider the detailed calculation in $4$-dimension. As suggested in the above section, the (approximate) expression for the entanglement entropy of the matter term depends on the position of the cutoff surface. In this section we will first consider the case that the cutoff surface is far from horizon.

 \renewcommand{\theequation}{3.\arabic{equation}}\setcounter{equation}{0}

  \subsection{With island}
  We consider one-sided black hole formed by spherical null shell collapsing at $v=v_0$. In the ``in'' region $v<v_0$, we have Minkowski metric
        \begin{eqnarray}\lb{Minkowski}
  ds^2=-du_{\text{in}} dv+r_{\text{in}}^2 d\Omega^2,
        \end{eqnarray}
        while in the ``out'' region $v>v_0$, we have Schwarzschild metric
         \begin{eqnarray}\lb{Sch}
  ds^2=-\left(1-\frac{r_h}{r}\right)du dv+r^2 d\Omega^2,
        \end{eqnarray}
   where
\begin{eqnarray}
 r_h=2 G_N M,\; u_{\text{in}}=t_{\text{in}}-r_{\text{in}},\;  u=t-r^{*}, \; v=t_{\text{in}}+r_{\text{in}}=t+r^{*}, \; r^{*}=r+r_h \ln \frac{|r-r_h|}{r_h}.
      \end{eqnarray}
To have smooth metric at the null shell, we have connecting condition \cite{Fabbri:2005mw}
    \begin{eqnarray}
  u=u_{\text{in}}-2r_h\ln \left(\frac{v_0-2r_h-u_{\text{in}}}{2r_h}\right),
    \end{eqnarray}
    then
        \begin{eqnarray}
    u_{\text{in}}=v_0-2r_h-2r_h W[e^{-1+\frac{v_0}{2r_h}-\frac{u}{2r_h}}],
       \end{eqnarray}
  where $W[z]$ is Lambert W function, also called product logarithm.

 Since $W[z] \simeq z$ when $z \rightarrow 0$, we have at late times $u \rightarrow + \infty$
  \begin{eqnarray}
 u_{\text{in}}(u) \simeq -2 r_h e^{-\frac{u}{2 r_h}}-2 r_h+v_0.
  \end{eqnarray}
  Notice that this is only valid outside horizon.

On the other hand, in terms of the Kruskal coordinates
    \begin{eqnarray}\lb{Kruskal-out}
    U&=&-2 r_h e^{-\frac{u}{2 r_h}}, \; V=2 r_h e^{\frac{v}{2 r_h}}, \; \text{(Outside horizon)}\\
        U&=&2 r_h e^{-\frac{(t-r^{*})}{2 r_h}}, \; V=2 r_h e^{\frac{(t+r^{*})}{2 r_h}}, \; \text{(Inside horizon)}\lb{Kruskal-in}
    \end{eqnarray}
 \eqref{Sch} is converted to
        \begin{eqnarray}
      ds^2&=&-e^{2\rho(U,V)}dU dV+r^2 d\Omega^2\\
     &=& -\frac{r_h e^{-r/(r_h)}}{r}dU dV+r^2 d\Omega^2.
        \end{eqnarray}
    Then at late times
    \begin{eqnarray}\lb{uin-U}
     u_{\text{in}} \simeq v_H+U,
    \end{eqnarray}
    where $v_H=v_0-2r_h$.
    Since both $U$ and $u_{\text{in}}$ are smoothly defined inside and outside horizon, eq.(\ref{uin-U}) is also valid inside horizon (at least in the vicinity of horizon),\footnote{Actually, for the case inside horizon (i.e. $u_{\text{in}}>v_0-2r_h$), we have connecting condition $  u=u_{\text{in}}-2r_h\ln \left(\frac{u_{\text{in}}-v_0+2r_h}{2r_h}\right)$ instead (here we still define $u=t-r^*$ inside horizon), then at late times, we have $ u_{\text{in}}(u) \simeq 2 r_h e^{-\frac{u}{2 r_h}}-2 r_h+v_0$. By using \eqref{Kruskal-in}, we still obtain the same expression \eqref{uin-U} for the case inside horizon.} 
     then
        \begin{eqnarray}\lb{derivative}
    \frac{dU}{d u_{\text{in}}} \simeq 1, \; \frac{dV}{d v}=\frac{V}{2 r_h}.
        \end{eqnarray}
        Now $ds^2=-e^{2\rho(u_{\text{in}},v)}du_{\text{in}} dv$ covers the whole spacetimes. In ``in'' region $v<v_0$, $e^{2\rho(u_{\text{in}},v)}=1$, and in ``out'' region, $e^{2\rho(u_{\text{in}},v)}$ is related to conformal factor $e^{2\rho(U,V)}$ in Kruskal coordinates via coordinate transformation
        \bqn\lb{co-trans}
        e^{2\rho(u_{\text{in}},v)}&=&e^{2\rho(U,V)}\frac{dU}{d u_{\text{in}}}\frac{dV}{d v}\\
&\approx&\frac{r_h e^{-r/r_h}}{r}\frac{V}{2 r_h},
        \eqn
        where we have used eq.(\ref{derivative}).

   Now let us turn to generalized entropy of the system, which is given by
  \bqn\lb{sgen-in}
S_{\text{gen}}=S_{\text{gravity}}+S_{\text{ent}},
\eqn
 where
\bqn\lb{grav}
S_{\text{gravity}}=\frac{\pi r_I^2}{G_N}
\eqn
is given by the area of boundary of island $\partial I$.

Since we are considering the case that cutoff surface $A$ is far from horizon, thus s-wave approximation is valid, then matter part is given by eq.(\ref{sent}). As explained previously, now we need to consider ``in'' vacuum state $|in\rangle$ of Minkowski region $v<v_0$. $|in\rangle$ is defined with respect to coordinates $(v,u_{\text{in}})$ (i.e. with respect to the modes $(4 \pi \omega)^{-1/2} e^{-i\omega v}$ and $(4 \pi \omega)^{-1/2} e^{-i\omega u_{\text{in}}}$), thus VEV of normal ordered stress tensor in coordinates $(v,u_{\text{in}})$ vanishes $\langle in|:T_{u_{\text{in}} u_{\text{in}}}:|in\rangle=\langle in|:T_{v v}:|in\rangle=0$ \cite{Fabbri:2005mw}.  Then $t_{\pm}=0$ for vacuum state $|vac\rangle=|in\rangle$ in $(v,u_{\text{in}})$ coordinates, thus
\begin{eqnarray}\lb{eq2.18}
d(A,I)=\sqrt{[u_{\text{in}}(A)-u_{\text{in}}(I)][v_I-v_A]}
\end{eqnarray}
is the distance between $A$ and $\partial I$ in flat metric $ds^2=-du_{\text{in}} dv$ and
\begin{eqnarray}\lb{eq2.19}
(e^{2\rho})_{t_{\pm}=0}=e^{2\rho(u_{\text{in}},v)}.
\end{eqnarray}

Actually, up to now, almost in all papers that discuss island in Schwarzschild spacetimes \cite{Alishahiha:2020qza,Hashimoto:2020cas,Matsuo:2020ypv,Arefeva:2021kfx} or Reissner-Nordstr\"om black hole \cite{Kim:2021gzd,Wang:2021woy}, Hartle-Hawking state $|H\rangle$ \cite{Hartle:1976tp} is taken into account.\footnote{While in \cite{Dong:2020uxp}, the authors considered eternal Schwarzschild black hole but with states defined with respect to $(\sinh^{-1} V, \sinh^{-1} U)$.} $|H\rangle$ is defined with respect to Kruskal coordinates $(V,U)$ (i.e. with respect to the modes $(4 \pi \omega)^{-1/2} e^{-i\omega V}$ and $(4 \pi \omega)^{-1/2} e^{-i\omega U}$) \cite{Fabbri:2005mw},
\bqn
\langle H|:T_{UU}: |H\rangle=\langle H|:T_{VV}: |H\rangle=0.
\eqn
Then for eq.(\ref{sent}),
\bqn\lb{HHd}
d(A,I)=\sqrt{(U(I)-U(A))(V(A)-V(I))},
\eqn
and
\bqn\lb{HHrho}
(e^{2\rho})_{t_{\pm}=0}=e^{2\rho(U,V)}.
\eqn
In \cite{Alishahiha:2020qza,Hashimoto:2020cas,Matsuo:2020ypv,Arefeva:2021kfx,Kim:2021gzd,Wang:2021woy}, the authors used eqs.(\ref{HHd}) and (\ref{HHrho}). This is the key difference between our paper and the others.

Put eqs.(\ref{sent}), (\ref{co-trans}), (\ref{grav}), (\ref{eq2.18}) and (\ref{eq2.19}) into eq.(\ref{sgen-in}), we have
\begin{eqnarray}
S_{\text{gen}}&=&\frac{\pi r_I^2}{G_N} +\frac{c}{12}  \ln \left(
[v_H+U_A-(v_H+U_I)]^2[v_I-v_A]^2 \frac{ e^{-(r_A+r_I)/r_h}V_AV_I}{4r_Ar_I}
\right)\\
\lb{secondline}
&=&\frac{c}{12}  \ln \left(\frac{V_A V_I \left(U_A-U_I\right)^2 e^{\chi_I} \left(\ln \left(\frac{V_I}{V_A}\right)\right){}^2}{e^2\left(1-\chi_I\right) \left(e^{W\left(-\chi_A\right)}-\chi_A\right)}\right)+\frac{\pi r_h^2  \left(1-\chi_I\right)^2}{G_N},
\end{eqnarray}
where $\chi_i\equiv \frac{U_iV_i}{4er_h^2}$ ($i=A, I$).
To obtain the second line of \eqref{secondline} we also have used
\bqn
r_A &=&r_h \left(1+W\left[e^{\frac{r_A^*}{r_h}-1}\right] \right),\\
r_I &\approx& r_h \left(1-e^{\frac{r_I^*}{r_h}-1} \right),\lb{r-rs}
\eqn
where $r_A^*$, $r_I^*$ can be converted to Kruskal coordinates by (\ref{Kruskal-out}) and (\ref{Kruskal-in}) respectively.\footnote{In (\ref{r-rs}), we implicitly assume $r_I<r_h$. For the case $r_I>r_h$, we can use $r_I \approx r_h \left(1+e^{\frac{r_I^*}{r_h}-1} \right)$ and $r_I^*$ can be converted to Kruskal coordinates by eq.(\ref{Kruskal-out}), then we will obtain same expression of $S_{\text{gen}}$ as (\ref{secondline}). Thus calculation in this subsection is valid both for the cases that $\partial I$ is inside or outside horizon. We will finally find that $\partial I$ is inside horizon when cutoff surface is far from horizon. }

We assume $\partial I$ is near horizon (which will be confirmed in eq.(\ref{eq2.29})), then $r^{*}_I \rightarrow -\infty$ thus $U_I \rightarrow 0$. Expand $S_{\text{gen}}$ to first order of $U_I$, we have
\begin{eqnarray}\lb{entropy}
S_{\text{gen}}&=&\left(\frac{\pi  r_h^2}{G_N}+\frac{1}{12} c \ln \left(\frac{4 V_I U_A \chi_A r_h^2 \ln ^2\left(\frac{V_I}{V_A}\right)}{ e \left( e^{W\left(-\chi_A\right)}-\chi_A\right)}\right)\right)\nb\\
&&+U_I \left(\frac{1}{12} c \left(-\frac{2}{U_A}+\frac{V_I}{2 e r_h^2}\right)-\frac{\pi  V_I}{2 e G_N}\right)+O\left(U_I^2\right).
\end{eqnarray}
  Extremizing (\ref{entropy}) over $U_I$, we get
 \begin{eqnarray}
 \frac{\partial S_{\text{gen}}}{\partial U_I}\approx -\frac{c \left(4 e r_h^2-V_I U_A\right)}{24 e U_A r_h^2}-\frac{\pi  V_I}{2 e G_N}=0,\nb\\
 \end{eqnarray}
 which has solution
  \begin{eqnarray}\lb{s1}
V_I= \frac{4 e c G_N r_h^2}{c G_N U_A-12 \pi  U_A r_h^2}.
   \end{eqnarray}
    Extremizing (\ref{entropy}) over $V_I$, we get
 \begin{eqnarray}
 \frac{\partial S_{\text{gen}}}{\partial V_I}\approx \frac{1}{24} \left(\frac{2 c \left(1+\frac{2}{\ln \left(\frac{V_I}{V_A}\right)}\right)}{V_I}+\frac{U_I \left(\frac{c}{r_h^2}-\frac{12 \pi }{G_N}\right)}{e}\right)=0.
 \end{eqnarray}
Dropping $\frac{2}{\ln \left(\frac{V_I}{V_A}\right)}$ due to its smallness, we have
 \begin{eqnarray}
\frac{U_I \left(\frac{c}{r_h^2}-\frac{12 \pi }{G_N}\right)}{e}+\frac{2 c}{V_I}=0,
 \end{eqnarray}
  which gives
  \begin{eqnarray}\lb{s2}
 U_I=-\frac{2 e c G_N r_h^2}{V_I \left(c G_N -12 \pi  r_h^2\right)}.
    \end{eqnarray}

 Eqs.(\ref{s1}) and (\ref{s2}) have solutions
   \begin{eqnarray}\lb{eq2.29}
U_I&=&-\frac{U_A}{2},\\
V_I&=&\frac{4 e c G_N r_h^2}{c G_N U_A-12 \pi  U_A r_h^2}.\lb{eq2.29-2}
    \end{eqnarray}
  Since $U_A<0$, thus $U_I>0$, which means $\partial I$ is inside horizon. Moreover,
 \bqn\lb{uivi}
\frac{ U_I V_I}{4r_h^2}=\frac{e c G_N}{24 \pi r_h^2-2c G_N} \approx \frac{e c G_N}{24 \pi r_h^2} \ll 1,
 \eqn
where we have used $c G_N \ll r_h^2$. Comparing (\ref{uivi}) with $\frac{ U_I V_I}{4r_h^2}=e^{r_I^*/r_h}$, we have $r_I^* \rightarrow -\infty$,    
     which means $\partial I$ is near horizon and this confirms our assumption.

Plug these solutions back to eq.(\ref{entropy}), we have
   \begin{eqnarray}
   S_{\text{gen}}&=&\frac{\pi  r_h^2}{G_N}+\frac{1}{12} c \left(\ln \left(-\frac{3 c G_N \chi_A  r_h^2 \ln ^2\left(-\frac{e c G_N}{3 \pi  U_A V_A}\right)}{\pi  \left( e^{W\left(-\chi_A\right)}-\chi_A \right)}\right)-1\right)+O\left(G_N^1\right)\\
 &\approx& \frac{\pi  r_h^2}{G_N}=S_{\text{BH}},\lb{island}
       \end{eqnarray}
 where in the last line we have taken into account that $c G_N \ll r_h^2$. Thus with island configuration, at late time, entropy of Hawking radiation is bounded by black hole Bekenstein-Hawking entropy which decreases monotonically due to black hole evaporation.

   \subsection{Without island}
 If there is no island, we can equivalently fix $r_I=0$ and $t_I$ as some constant. Then the area term (\ref{grav}) vanishes. We only have field term eq.(\ref{sent}), and $u_{\text{in}}(I)=t_I$, $e^{2\rho(u_{\text{in}},v)(I)}=1$ since now $r_I=0$ is in the Minkowski region (\ref{Minkowski}). As a consequence we have
 \begin{eqnarray}
S_{\text{gen}}=S_{\text{ent}}&=&\frac{c}{12}  \ln \left(
[v_H+U_A-t_I]^2[v_I-v_A]^2 \frac{r_h e^{-r_A/r_h}}{r_A}\frac{V_A}{2 r_h}
\right)\\
&=&\frac{1}{12} c \ln \left(\frac{r_h}{r_A} \sqrt{\frac{r_A}{r_h}-1} e^{-\frac{r_A+t_A}{2 r_h}} \left(2 e^{\frac{r_A}{2 r_h}} \sqrt{r_h \left(r_A-r_h\right)}+e^{\frac{t_A}{2 r_h}} \left(2 r_h+t_I-v_0\right)\right)^2\right)\nb\\
&+&\frac{1}{6} c \ln \left( \left(r_h \ln \left(\frac{r_A}{r_h}-1\right)+r_A+t_A-t_I\right)\right)\\
\lb{no-island}
&\approx&\frac{c}{24r_h} t_A,
        \end{eqnarray}
 where in the last line we have taken the late time limit $t_A \gg r_A (\gg r_h)$.\footnote{There is also a logarithm divergent term proportional to $\ln(t_A)$ in eq.(\ref{no-island}), but it is only subdominant, so we omitted it.} In a word, without island configuration, at late time the radiation entropy grows linearly with time, which is consistent with Hawking's result.

 Compare eq.(\ref{no-island}) with eq.(\ref{island}), we have
  \begin{eqnarray}
S(R)=\text{min}( S_{\text{gen}}) \approx \text{min}(S_{\text{ent}},S_{\text{BH}}),
         \end{eqnarray}
 thus the Page time is at
   \begin{eqnarray}\lb{Page}
 t_{\text{Page}}&\approx&\frac{24 r_h}{c}S_{\text{BH}}.
         \end{eqnarray}
 Our result for Page time is about twice as much as the Page time calculated in \cite{Alishahiha:2020qza}. This is due to the fact that we choose ``in'' vacuum state  $|in\rangle$ which has $t_{\pm}=0$ in $(v,u_{\text{in}})$ coordinate. While in \cite{Alishahiha:2020qza}, although the authors also considered dynamical black hole, they still started from Hartle-Hawking state  $|H\rangle$ which has $t_{\pm}=0$ in $(V,U)$ coordinate and $|H\rangle$ is thermal with respect to $(v,u_{\text{in}})$ coordinate. $S_{\text{ent}}$ of vacuum state  $|in\rangle$ is smaller than $S_{\text{ent}}$ of thermal state  $|H\rangle$, thus we will arrive at Page time later.

\section{Cutoff surface near horizon}\lb{4d-near}
 \renewcommand{\theequation}{4.\arabic{equation}}\setcounter{equation}{0}

  \subsection{With island}

In this subsection, we consider the case in which cutoff surface is near horizon and we will follow similar logic in \cite{Hashimoto:2020cas}. The authors in \cite{Hashimoto:2020cas} considered Hartle-Hawking state in eternal Schwarzschild black hole while we consider ``in'' vacuum state in one-sided black hole formed by collapse of null shell. Despite this fact, when length scale of the region is sufficiently small compared with length scale of the curvature, entropy of matter fields can be approximately given by the one in vacuum state in flat spacetime no matter which spacetime and state we consider \cite{Hashimoto:2020cas}.

 In this case we still have
 \bqn
S_{\text{gravity}}=\frac{\pi r_I^2}{G_N}.
\eqn
But entropy for fields is now given by eq.(\ref{sent2}).\footnote{Since eq.(\ref{sent2}) is valid for any state, we can set $x^+=V, x^-=U$ for simplicity.}

 Since we assume both $\partial I$ and cutoff surface is near horizon, $L$ can be approximately given by \cite{Hashimoto:2020cas}
\bqn
L&\approx&\sqrt{d(A,I)^2 e^{\rho_A} e^{\rho_I}}\nb\\
&=&\sqrt{\frac{r_h \left(U_A-U_I\right) \left(-V_A+V_I\right) e^{-\frac{r_A}{2 r_h}-\frac{r_I}{2 r_h}}}{\sqrt{r_I r_A}}}\nb\\
&=&\sqrt{\frac{4 r_h^3 e^{-\frac{r_A+t_A+r_I+t_I}{2 r_h}} \left(\sqrt{\frac{r_A}{r_h}-1} e^{\frac{r_A+t_I}{2 r_h}}-\sqrt{-1+\frac{r_I}{r_h}} e^{\frac{t_A+r_I}{2 r_h}}\right) \left(\sqrt{\frac{r_A}{r_h}-1} e^{\frac{r_A+t_A}{2 r_h}}-\sqrt{-1+\frac{r_I}{r_h}} e^{\frac{r_I+t_I}{2 r_h}}\right)}{\sqrt{r_I r_A}}}.\nb\\
 \eqn

Then we have
\bqn
S_{\text{gen}}&=&\frac{\pi r_I^2}{G_N}-\kappa c \frac{\text{Area}}{L^2}\\
&=&\frac{\pi  c \kappa  \sqrt{r_I r_A} r_A^2 e^{\frac{r_A+t_A+r_I+t_I}{2 r_h}}}{r_h^3 \left(\sqrt{\frac{r_A}{r_h}-1} e^{\frac{r_A+t_I}{2 r_h}}-\sqrt{-1+\frac{r_I}{r_h}} e^{\frac{t_A+r_I}{2 r_h}}\right) \left(-\sqrt{\frac{r_A}{r_h}-1} e^{\frac{r_A+t_A}{2 r_h}}+\sqrt{-1+\frac{r_I}{r_h}} e^{\frac{r_I+t_I}{2 r_h}}\right)}+\frac{\pi  r_I^2}{G_N}.\nb\\
\eqn
 Since we assume both $\partial I$ and cutoff surface are near and outside horizon\footnote{When cutoff surface is near horizon, there is no physical solution of $\partial I$ to be inside horizon, see Appendix \ref{app-a}.}, we can make replacement $r_A=r_h(1+\alpha)$ and $r_I=r_h(1+\beta)$, where $\beta < \alpha \ll 1$,
 then we have
 \bqn\lb{sgen-out}
S_{\text{gen}}= \frac{\pi  (\beta +1)^2 r_h^2}{G_N}-\frac{\pi  (\alpha +1)^{5/2} \sqrt{\beta +1} c \kappa }{e^{-\frac{\alpha }{2}-\frac{\beta }{2}} \left(e^{\alpha } \alpha +e^{\beta } \beta \right)-2 \sqrt{\alpha  \beta } \cosh \left(\frac{t_A-t_I}{2 r_h}\right)}.
 \eqn
 Extremizing (\ref{sgen-out}) over $t_I$, we get
 \bqn
  \frac{\partial S_{\text{gen}}}{\partial t_I}=\frac{\pi  (\alpha +1)^{5/2} \sqrt{\beta +1} c \kappa  e^{\alpha +\beta } \sqrt{\alpha  \beta } \sinh \left(\frac{t_A-t_I}{2 r_h}\right)}{r_h \left(e^{\alpha } \alpha -2 e^{\frac{\alpha +\beta }{2}} \sqrt{\alpha  \beta } \cosh \left(\frac{t_A-t_I}{2 r_h}\right)+e^{\beta } \beta \right)^2}=0,
  \eqn
  which has solutions
  \bqn
  t_I=t_A-4 i \pi  c_1 r_h,c_1\in \mathbb{Z},
  \eqn
  or
   \bqn
  t_I=t_A -2 (2 i \pi  c_1+i \pi ) r_h,c_1\in \mathbb{Z},
  \eqn
  where $\mathbb{Z}$ represents integers. Dropping complex solutions, we are left with
 \bqn
  t_I=t_A.
\eqn
This confirms the assumption $t_I=t_A$ in \cite{Hashimoto:2020cas}. Plug $t_I=t_A$ back into eq.(\ref{sgen-out}), and expand to first order of $\beta$, we have
   \bqn\lb{sgen-out2}
S_{\text{gen}}&&= \left(\frac{\pi  r_h^2}{G_N}-\frac{\pi  e^{-\alpha /2} (\alpha +1)^{5/2} c \kappa }{\alpha }\right)-\frac{2 \sqrt{\beta } \left(\pi  e^{-\alpha } (\alpha +1)^{5/2} c \kappa \right)}{\alpha ^{3/2}}\nb\\
&&+\beta  \left(\frac{2 \pi  r_h^2}{G_N}-\frac{\pi  e^{-3 \alpha /2} (\alpha +1)^{5/2} \left(e^{\alpha } \alpha +3\right) c \kappa }{\alpha ^2}\right)+O\left(\beta ^{3/2}\right).
\eqn
   Extremizing (\ref{sgen-out2}) over $\beta$, we get
 \bqn
  \frac{\partial S_{\text{gen}}}{\partial \beta} \approx \pi  \left(-\frac{e^{-\alpha } (\alpha +1)^{5/2} c \kappa }{\sqrt{\alpha ^3 \beta }}-\frac{e^{-\frac{3 \alpha }{2}} \left(e^{\alpha } \alpha +3\right) (\alpha +1)^{5/2} c \kappa }{\alpha ^2}+\frac{2 r_h^2}{G_N}\right)=0,
  \eqn
  which has solution
  \bqn\lb{beta}
\beta&&=  \frac{e^{\alpha } \alpha  (\alpha +1)^5 c^2 G_N^2 \kappa ^2}{\left(3 (\alpha +1)^{5/2} c G_N \kappa +e^{\alpha } \alpha  (\alpha +1)^{5/2} c G_N \kappa -2 e^{\frac{3 \alpha }{2}} \alpha ^2 r_h^2\right)^2}\\
&&\approx \frac{ c^2 G_N^2 \kappa ^2}{4 \alpha^3 r_h^4}=\frac{ c^2 G_N^2 \kappa ^2}{4 (r_A-r_h)^3 r_h}.\lb{beta2}
  \eqn
This matches the result in \cite{Hashimoto:2020cas} and confirms the assumption that $\partial I$ is near and outside horizon due to the fact that $c^2 G_N^2 \ll r_h^4$.
  Plug eq.(\ref{beta}) into (\ref{sgen-out2}) and expand the result to zeroth order in $G_N$, we have
     \bqn
S_{\text{gen}}&=&\frac{\pi  r_h^2}{G_N}-\frac{\pi  e^{-\alpha /2} (\alpha +1)^{5/2} c \kappa }{\alpha }+O\left(G_N^1\right)\\
 &\approx&\frac{\pi  r_h^2}{G_N}-\frac{\pi   c \kappa }{\alpha }+O\left(G_N^1\right)\\
 &=& \frac{\pi  r_h^2}{G_N}-\pi  c \kappa \frac{r_h }{(r_A-r_h) }+O\left(G_N^1\right)\\
 &\approx&  \frac{\pi  r_h^2}{G_N}=S_{\text{BH}},\lb{island2}
  \eqn
  where in the last line we have taken into account that $c G_N \ll r_h^{2}$. Thus for the case that cutoff surface near horizon, with island configuration, at late time, entropy of Hawking radiation is bounded by black hole Bekenstein-Hawking entropy which decreases monotonically due to black hole evaporation.
    \subsection{Without island}
  In the case without island, we still need to use eq.(\ref{sent}) to calculate $S_{\text{ent}}$ due to the fact that $r_I=0$ is far from near-horizon cutoff surface for macroscopic black hole ($r_h \gg \ell_p^2$). Then the result of $S_{\text{gen}}$ without island for near-horizon cutoff surface will be same as eq.(\ref{no-island}). Comparing with (\ref{island2}), we find same Page time
     \begin{eqnarray}\lb{Page2}
 t_{\text{Page}}&\approx&\frac{24 r_h}{c}S_{\text{BH}}.
         \end{eqnarray}

\section{Higher dimensions}\lb{higher-d}
 \renewcommand{\theequation}{5.\arabic{equation}}\setcounter{equation}{0}

To examine our result obtained in previous sections more comprehensively,  in this section, we will consider one-sided black hole in $D$ ($D>4$) dimensional spacetime formed by null shell collapsing at $v=v_0$.

Similarly, in the ``in'' region $v<v_0$, we have Minkowski metric
        \begin{eqnarray}\lb{Minkowski-d}
  ds^2=-du_{\text{in}} dv+r_{\text{in}}^2 d\Omega^2_{D-2},
        \end{eqnarray}
        and in the ``out'' region $v>v_0$, we have Schwarzschild metric
         \begin{eqnarray}\lb{Schwarzschild-d}
  ds^2=-\left(1-\frac{r_h^{D-3}}{r^{D-3}}\right)du dv+r^2 d\Omega^2_{D-2}.
        \end{eqnarray}
We have defined $r_h$ in higher dimensions as\footnote{$\Omega_{D-2}=\frac{2 \pi^{\frac{D-1}{2}}}{\Gamma(\frac{D-1}{2})}$ is the area of a unit $D-2$ dimensional sphere $\mathbf S^{D-2}$.}
\begin{eqnarray}\lb{uv-D}
 r_h&=&\left(\frac{16 \pi G_N^{(D)} M}{(D-2) \Omega_{D-2}} \right)^{1/(D-3)}, \; u_{\text{in}}=t_{\text{in}}-r_{\text{in}},\;  u=t-r^{*}, \; v=t_{\text{in}}+r_{\text{in}}=t+r^{*}, \\
   r^{*}&=&\Pr \int_0^r \frac{dr'}{(1-\frac{r_h^{D-3}}{r'^{D-3}})}=\begin{cases}
 r^{*}_{<} \equiv \frac{-r^{D-2}}{(D-2)r_h^{D-3}} {}_{2}{F}_1 (1;1+\frac{1}{D-3},2+\frac{1}{D-3};(\frac{r}{r_h})^{D-3}), \; r<r_h,\\
 r^{*}_{>} \equiv    \frac{-r^{D-2}}{(D-2)r_h^{D-3}} {}_{2}{F}_1 (1;1+\frac{1}{D-3},2+\frac{1}{D-3};(\frac{r}{r_h})^{D-3})-\frac{r_h \ln(-1)}{D-3}, \; r>r_h,
   \end{cases}
      \end{eqnarray}
where $\Pr$ means principle value integral.

 Now the Kruskal coordinates are
    \begin{eqnarray}\lb{Kruskal-out-d}
    U&=&-2 r_h e^{-(D-3)\frac{u}{2 r_h}}, \; V=2 r_h e^{(D-3)\frac{v}{2 r_h}}, \; \text{(Outside horizon)}\\
        U&=&2 r_h e^{-(D-3)\frac{(t-r^{*})}{2 r_h}}, \; V=2 r_h e^{(D-3)\frac{(t+r^{*})}{2 r_h}}, \; \text{(Inside horizon)}\lb{Kruskal-in-d}
    \end{eqnarray}
    and (\ref{Schwarzschild-d}) is converted to
        \begin{eqnarray}
      ds^2&=&-e^{2\rho(U,V)}dU dV+r^2 d\Omega^2_{D-2}\\
     &=& -\frac{\left(1-\frac{r_h^{D-3}}{r^{D-3}}\right)}{(D-3)^2 \exp((D-3)\frac{r^{*}_{>}}{r_h}) }dU dV+r^2 d\Omega^2_{D-2}.
        \end{eqnarray}

                  \subsection{Connecting condition}

   To have smooth metric at the null shell, metric on both sides of null shell needs to be the same. Thus we have connection condition \cite{Fabbri:2005mw}
\bqn\lb{eq4.9}
r_{\text{in}}(v_0,u_{\text{in}})=r(v_0,u),
\eqn
where
\bqn\lb{eq4.10}
r_{\text{in}}(v_0,u_{\text{in}})=\frac{v_0-u_{\text{in}}}{2},
\eqn
and $r(v_0,u)$ is given implicitly outside horizon by
\bqn
\frac{v_0-u}{2}= r^{*}_{>}(v_0,u) \equiv  \frac{-r(v_0,u)^{D-2}}{(D-2)r_h^{D-3}} {}_{2}{F}_1 (1;1+\frac{1}{D-3},2+\frac{1}{D-3};(\frac{r(v_0,u)}{r_h})^{D-3})-\frac{r_h \ln(-1)}{D-3}, \; r>r_h.\nb\\
\eqn
Thus we have connection condition outside horizon
   \begin{eqnarray}\lb{eq4.12}
u=\frac{2 r(v_0,u)^{D-2} r_h^{3-D} \, _2F_1\left(1,1+\frac{1}{D-3};2+\frac{1}{D-3};\left(\frac{r(v_0,u)}{r_h}\right){}^{D-3}\right)}{D-2}+\frac{2 i \pi  r_h}{D-3}+v_0.
          \end{eqnarray}
At late times, $u \rightarrow +\infty$, $u_{\text{in}} \rightarrow v_0-2 r_h$. Plugging $r(v_0,u)=\frac{v_0-u_{\text{in}}}{2}$ into eq.(\ref{eq4.12}) and expanding it around $u_{\text{in}} = v_0-2 r_h$, we get\footnote{$\gamma \approx 0.577216$ is Euler's constant and $\psi ^{(n)} (z)$ the $n^{th}$ derivative of the digamma function $\psi (z)$.}
      \begin{eqnarray}
   u&=&\frac{2^{3-D} r_h^{3-D} \left(v_0-u_{\text{in}}\right){}^{D-2} \, _2F_1\left(1,1+\frac{1}{D-3};2+\frac{1}{D-3};2^{3-D} \left(\frac{v_0-u_{\text{in}}}{r_h}\right){}^{D-3}\right)}{D-2}+\frac{2 i \pi  r_h}{D-3}+v_0\\
   &\approx& \frac{-2 r_h \ln \left(\frac{D-3}{2 r_h}\right)-2 \psi ^{(0)}\left(1+\frac{1}{D-3}\right) r_h+D v_0-2 r_h \ln \left(2 r_h+u_{\text{in}}-v_0\right)+2 i \pi  r_h-2 \gamma  r_h-3 v_0}{D-3}\nb\\
   &&+O\left(\left(2 r_h+u_{\text{in}}-v_0\right)^1\right) \lb{eq5.14}.
             \end{eqnarray}
From \eqref{eq5.14} we have
          \begin{eqnarray}
 u_{\text{in}}  \approx \exp \left(\frac{-2 r_h \ln \left(\frac{D-3}{2 r_h}\right)-2 \psi ^{(0)}\left(1+\frac{1}{D-3}\right) r_h-D u+D v_0+2 i \pi  r_h-2 \gamma  r_h+3 u-3 v_0}{2 r_h}\right)-2 r_h+v_0.\nb\\
                \end{eqnarray}
However, the above formula is only valid outside horizon. This is because outside horizon we have
  \begin{eqnarray}
   u=\frac{2 r_h}{-(D-3)} \ln \frac{U}{-2 r_h}.
   \end{eqnarray}
   Then\footnote{In $D \rightarrow 4$ limit, eq.(\ref{uin-U-d-far}) has additional proportional constant when compared with (\ref{uin-U}). This is because for (\ref{uin-U-d-far}) we take late time limit first and then solve for $u_{\text{in}}$, while for (\ref{uin-U}) we solve for $u_{\text{in}}$ first and then take late time limit. This will not affect the physical content.}
     \begin{eqnarray} \lb{uin-U-d-far}
 u_{\text{in}}  \approx   \frac{U e^{\frac{(D-3) v_0}{2 r_h}-\psi ^{(0)}\left(1+\frac{1}{D-3}\right)-\gamma }}{D-3}-2 r_h+v_0,
      \end{eqnarray}
which indicates that
       \begin{eqnarray}\lb{derivative-d1}
    \frac{dU}{d u_{\text{in}}} \simeq \frac{D-3}{e^{\frac{(D-3) v_0}{2 r_h}-\psi ^{(0)}\left(1+\frac{1}{D-3}\right)-\gamma }}, \; \frac{dV}{d v}=\frac{(D-3)V}{2 r_h},
        \end{eqnarray}
      and  thus
    \begin{eqnarray}
e^{2\rho(u_{\text{in}},v)}&=&e^{2\rho(U,V)}\frac{dU}{d u_{\text{in}}}\frac{dV}{d v}\\
&\approx& e^{2\rho(U,V)} \frac{D-3}{e^{\frac{(D-3) v_0}{2 r_h}-\psi ^{(0)}\left(1+\frac{1}{D-3}\right)-\gamma }} \frac{(D-3)V}{2 r_h}\\
&=&\frac{(1-\frac{r_h^{D-3}}{r^{D-3}})}{(D-3)^2 \exp((D-3)\frac{r^{*}_{>}}{r_h}) } \frac{D-3}{e^{\frac{(D-3) v_0}{2 r_h}-\psi ^{(0)}\left(1+\frac{1}{D-3}\right)-\gamma }} \frac{(D-3)V}{2 r_h}.\lb{out-far-d}
\end{eqnarray}
  Similar to the case for $4D$, since both $U$ and $u_{\text{in}}$ are smoothly defined inside and outside horizon, eq.(\ref{out-far-d}) is also valid inside horizon (at least in the vicinity of horizon). This is explicitly shown in Appendix. \ref{apx.b}.

\subsection{Cutoff surface far from horizon}
\subsubsection{with island}
Again the generalized entropy is given by
\begin{eqnarray}
S_{\text{gen}}=S_{\text{grav}} +S_{\text{ent}},
\end{eqnarray}
where area term is
\bqn\lb{grav-d}
S_{\text{grav}}=\frac{\Omega_{D-2} r_I^{D-2}}{4G_N^{(D)}},
\eqn
and based on similar logic of the case of $4D$, we can also make $s$-wave approximation for $S_{\text{ent}}$ in higher dimensions. Then matter term is
\bqn\lb{ent-d}
S_{\text{ent}}=\frac{c}{12}  \ln \left( d(A,I)^4 e^{2\rho_A} e^{2\rho_I}
\right)_{t_{\pm}=0},
\eqn
where $(e^{2\rho_A})_{t_{\pm}=0}$ should be in the form of eq.(\ref{out-far-d}) while $(e^{2\rho_I})_{t_{\pm}=0}$ should be in the form of eq.(\ref{in-near-d}). In what follows we will assume $c G_N^{(D)} \ll r_h^{D-2}$. Put all things together, we have
\begin{eqnarray}
S_{\text{gen}}=\frac{2 \pi^{\frac{D-1}{2}} r_I^{D-2}}{4\Gamma(\frac{D-1}{2})G_N^{(D)}}+\frac{c}{12}  \ln \left( (u_{\text{in}}(A)-u_{\text{in}}(I))^2 (v_I-v_A)^2 e^{2\rho_A} e^{2\rho_I}
\right)_{t_{\pm}=0}.
\end{eqnarray}
Next we need to convert all variables to Kruskal coordinates. For $r_A \gg r_h$,
  \bqn
   r^{*}_{>}(A) \equiv    \frac{-r_A^{D-2}}{(D-2)r_h^{D-3}} {}_{2}{F}_1 (1;1+\frac{1}{D-3},2+\frac{1}{D-3};(\frac{r_A}{r_h})^{D-3})-\frac{r_h \ln(-1)}{D-3}, \; r>r_h.
   \eqn
      Define a dimensionless constant $a=\left(\frac{r_A}{r_h}\right)^{D-3}\gg1$. We then expand $r^{*}_{>}(A)$ around $a=\infty$,
   \bqn
   r^{*}_{>}(A) &&= \frac{-(a^{\frac{1}{D-3}}r_h)^{D-2}}{(D-2)r_h^{D-3}} {}_{2}{F}_1 (1;1+\frac{1}{D-3},2+\frac{1}{D-3};a)-\frac{r_h \ln(-1)}{D-3}\\
   &&\approx a^{\frac{1}{D-3}} \left(r_h-\frac{r_h}{(D-4) a}+O\left(\left(\frac{1}{a}\right)^2\right)\right)+\frac{\left((-1)^{\frac{1}{3-D}} \Gamma \left(\frac{1}{3-D}\right) \Gamma \left(\frac{1}{D-3}\right)-i \pi  (D-3)\right) r_h}{(D-3)^2}\nb\\
   \\
   &&\approx a^{\frac{1}{D-3}} r_h\\
   &&=r_A.\lb{eq5.30}
   \eqn
Then $r^*_>(A)$ can be converted to Kruskal coordinates via eq.(\ref{uv-D}) and (\ref{Kruskal-out-d}).

  On the other hand, since $r_I$ is close to horizon, we can expand it around horizon, i,e., $r_I = r_h(1-\epsilon)$ with $\epsilon\ll1$, then
   \bqn
    r^{*}_{<}(I) &&\equiv \frac{-r_I^{D-2}}{(D-2)r_h^{D-3}} {}_{2}{F}_1 (1;1+\frac{1}{D-3},2+\frac{1}{D-3};(\frac{r_I}{r_h})^{D-3}), \; r<r_h,\\
    &&=\frac{-(r_h(1-\epsilon))^{D-2}}{(D-2)r_h^{D-3}} {}_{2}{F}_1 (1;1+\frac{1}{D-3},2+\frac{1}{D-3};(1-\epsilon)^{D-3}), \; r<r_h,\\
    &&\approx \frac{r_h \left(\ln (D-3)+\psi ^{(0)}\left(1+\frac{1}{D-3}\right)+\ln (\epsilon )+\gamma \right)}{D-3}+O\left(\epsilon \right)\\
        &&\approx \frac{r_h \left(\ln (D-3)+\psi ^{(0)}\left(1+\frac{1}{D-3}\right)+\ln (1-\frac{r_I}{r_h} )+\gamma \right)}{D-3}.
    \eqn
   Thus
   \bqn
 r_I \approx  r_h \left(1-\exp \left(\frac{-\ln (D-3) r_h-\psi ^{(0)}\left(1+\frac{1}{D-3}\right) r_h+D r^{*}_{<}(I)-\gamma  r_h-3 r^{*}_{<}(I)}{r_h}\right)\right),\lb{eq5.35}
   \eqn
   and now $r^{*}_{<}(I)$ can be converted to Kruskal coordinates via eq.(\ref{uv-D}) and (\ref{Kruskal-in-d}).\footnote{To deduce (\ref{eq5.35}), we implicitly assume $r_I<r_h$. For the case $r_I>r_h$, we can use $r_I \approx  r_h \left(1+\exp \left(\frac{-\ln (D-3) r_h-\psi ^{(0)}\left(1+\frac{1}{D-3}\right) r_h+D r^{*}_{>}(I)-\gamma  r_h-3 r^{*}_{>}(I)}{r_h}\right)\right)$ and $r_I^*$ can be converted to Kruskal coordinates by eq.(\ref{Kruskal-out-d}), then we will obtain same expression of $S_{\text{gen}}$ as (\ref{gen-d}). Thus calculation in this subsection is valid both for the cases that $\partial I$ is inside or outside horizon. We will finally find that $\partial I$ is inside horizon when cutoff surface is far from horizon. }

   Finally, we have\footnote{In $D \rightarrow 4$ limit, eq.(\ref{gen-d}) is slightly different from eq.(\ref{secondline}). This is because that we take approximation (\ref{eq5.30}) and the fact that in $D \rightarrow 4$ limit, eq.(\ref{uin-U-d-far}) has additional proportional constant when compared with (\ref{uin-U}). But this will not change the physical content.}
   \bqn\lb{gen-d}
   S_{\text{gen}}&&\approx \frac{2^{3-2 D} \pi ^{\frac{D-1}{2}} \left(\frac{r_h \left(-\frac{U_I V_I e^{-\psi ^{(0)}\left(1+\frac{1}{D-3}\right)-\gamma }}{r_h^2}+4 D-12\right)}{D-3}\right)^{D-2}}{G_N^{(D)} \Gamma \left(\frac{D-1}{2}\right)}\nb\\
   &&+\frac{1}{12} c \ln \left(\ln ^2\left(\frac{V_A}{V_I}\right) \left(1-(D-3)^{D-3} \ln ^{3-D}\left(-\frac{U_A V_A}{4 r_h^2}\right)\right)\right)\nb\\
   &&+\frac{1}{12} c \ln \left(16 \left(U_A-U_I\right){}^2 r_h^4 \left(1-r_h^{D-3} \left(r_h-\frac{U_I V_I e^{-\psi ^{(0)}\left(1+\frac{1}{D-3}\right)-\gamma }}{4 (D-3) r_h}\right)^{3-D}\right)\right)\nb\\
   &&-\frac{1}{12} c \ln \left((D-3)^4 U_I U_A\right).
   \eqn

 We assume $\partial I$ is near horizon, and expand $S_{\text{gen}}$ to first order in $U_I$, then extremize the result over $U_I$, we get
 \begin{eqnarray}
 \frac{\partial S_{\text{gen}}}{\partial U_I} \approx \frac{1}{96} c \left(-\frac{16}{U_A}+\frac{(D-2) V_I e^{-\psi ^{(0)}\left(1+\frac{1}{D-3}\right)-\gamma }}{(D-3) r_h^2}\right)-\frac{(D-2) \pi ^{\frac{D-1}{2}} V_I e^{-\psi ^{(0)}\left(1+\frac{1}{D-3}\right)-\gamma } r_h^{D-4}}{8 (D-3) G_N^{(D)} \Gamma \left(\frac{D-1}{2}\right)}=0,\nb\\
\eqn
thus
   \bqn\lb{vi-d}
V_I \approx   \frac{16 \sqrt{\pi } c (D-3) G_N^{(D)} \Gamma \left(\frac{D-1}{2}\right) e^{\psi ^{(0)}\left(1+\frac{1}{D-3}\right)+\gamma } r_h^4}{(D-2) U_A \left(\sqrt{\pi } c G_N^{(D)} \Gamma \left(\frac{D-1}{2}\right) r_h^2-12 \pi ^{D/2} r_h^D\right)}.
   \eqn
   On the other hand, since we assume $\partial I$ is near horizon, we can expand $S_{\text{gen}}$ to first order in $U_I$ and extremize the result over $V_I$, then we get
   \bqn
 &&\frac{\partial S_{\text{gen}}}{\partial V_I}  \nb\\
 &&\approx  \frac{e^{-\psi ^{(0)}\left(1+\frac{1}{D-3}\right)-\gamma }}{96 \sqrt{\pi } (D-3) G_N^{(D)} V_I \Gamma \left(\frac{D-1}{2}\right) r_h^4 \ln \left(\frac{V_A}{V_I}\right)} \Bigg(-12 (D-2) \pi ^{D/2} U_I V_I \ln \left(\frac{V_A}{V_I}\right) r_h^D\nb\\
 &&+\sqrt{\pi } c G_N^{(D)} \Gamma \left(\frac{D-1}{2}\right) r_h^2 \left(-16 (D-3) e^{\psi ^{(0)}\left(1+\frac{1}{D-3}\right)+\gamma } r_h^2+\ln \left(\frac{V_A}{V_I}\right) \left(8 (D-3) e^{\psi ^{(0)}\left(1+\frac{1}{D-3}\right)+\gamma } r_h^2+(D-2) U_I V_I\right)\right)
 \Bigg)\nb\\
 &&\approx \frac{e^{-\psi ^{(0)}\left(1+\frac{1}{D-3}\right)-\gamma } \left(\sqrt{\pi } c G_N^{(D)} \Gamma \left(\frac{D-1}{2}\right) r_h^2 \left(8 (D-3) e^{\psi ^{(0)}\left(1+\frac{1}{D-3}\right)+\gamma } r_h^2+(D-2) U_I V_I\right)-12 (D-2) \pi ^{D/2} U_I V_I r_h^D\right)}{96 \sqrt{\pi } (D-3) G_N^{(D)} V_I \Gamma \left(\frac{D-1}{2}\right) r_h^4}=0\nb\\
   \eqn
   where we have taken into account that $\ln \left(\frac{V_A}{V_I}\right) \gg 1$. Then
\bqn
U_I \approx -\frac{8 \sqrt{\pi } c (D-3) G_N^{(D)} \Gamma \left(\frac{D-1}{2}\right) e^{\psi ^{(0)}\left(1+\frac{1}{D-3}\right)+\gamma } r_h^4}{(D-2) V_I \left(\sqrt{\pi } c G_N^{(D)} \Gamma \left(\frac{D-1}{2}\right) r_h^2-12 \pi ^{D/2} r_h^D\right)},
\eqn
together with eq.(\ref{vi-d}), we have
\bqn\lb{ub-d-final}
U_I &&\approx -\frac{U_A}{2},\\
V_I &&\approx \frac{16 \sqrt{\pi } c (D-3) G_N^{(D)} \Gamma \left(\frac{D-1}{2}\right) e^{\psi ^{(0)}\left(1+\frac{1}{D-3}\right)+\gamma } r_h^4}{(D-2) U_A \left(\sqrt{\pi } c G_N^{(D)} \Gamma \left(\frac{D-1}{2}\right) r_h^2-12 \pi ^{D/2} r_h^D\right)}.\lb{ub-d-final-2}
\eqn
As in the case of $4D$, $U_A<0$, thus $U_I>0$, which means $\partial I$ is inside horizon. Moreover,
 \bqn\lb{uivi-d}
\frac{ U_I V_I}{4r_h^2}=-\frac{2 \sqrt{\pi } c (D-3) G_N^{(D)} \Gamma \left(\frac{D-1}{2}\right) e^{\psi ^{(0)}\left(1+\frac{1}{D-3}\right)+\gamma } r_h^2}{(D-2) \left(\sqrt{\pi } c G_N^{(D)} \Gamma \left(\frac{D-1}{2}\right) r_h^2-12 \pi ^{D/2} r_h^D\right)} \ll 1,
 \eqn
where we have used $c G_N^{(D)} \ll r_h^{D-2}$. Comparing (\ref{uivi-d}) with $\frac{ U_I V_I}{4r_h^2}=e^{(D-3)r_I^*/r_h}$, we have $r_I^* \rightarrow -\infty$,    which means $\partial I$ is near horizon and this confirms our assumption. And take into account that $c G_N^{D} \ll r_h^{D-2}$, in the $D \rightarrow 4$ limit, eq.(\ref{ub-d-final-2}) will reduce to the $4D$ result (\ref{eq2.29-2}).

Plug these solutions back to eq.(\ref{gen-d}) and expand the result to zeroth order in $G_N^{(D)}$, we have
   \begin{eqnarray}
   S_{\text{gen}}&=&\frac{\pi ^{\frac{D}{2}-\frac{1}{2}} r_h^{D-2}}{2 G_N^{(D)} \Gamma \left(\frac{D-1}{2}\right)}\nb\\
  &&+ \frac{c}{12} \Bigg(-1+\ln \Bigg(\frac{12 c \pi ^{\frac{1}{2}-\frac{D}{2}} G_N^{(D)} \Gamma \left(\frac{D-1}{2}\right)}{(D-3)^6 (D-2)}\left((D-3)^3 \ln ^D\left(-\frac{U_A V_A}{4 r_h^2}\right)-(D-3)^D \ln ^3\left(-\frac{U_A V_A}{4 r_h^2}\right)\right)\nb\\
  &&r_h^6 \left(r_h \ln \left(-\frac{U_A V_A}{4 r_h^2}\right)\right)^{-D} \ln ^2\left(-\frac{3 (D-2) \pi ^{\frac{D-1}{2}} U_A V_A e^{-\psi ^{(0)}\left(1+\frac{1}{D-3}\right)-\gamma } r_h^{D-4}}{4 c (D-3) G_N^{(D)} \Gamma \left(\frac{D-1}{2}\right)}\right)
  \Bigg)
  \Bigg)+\mathcal{O}(G_N^{(D)})\\
  &\approx& \frac{\pi ^{\frac{D}{2}-\frac{1}{2}} r_h^{D-2}}{2 G_N^{(D)} \Gamma \left(\frac{D-1}{2}\right)}=S_{\text{BH}},\lb{island-d}
       \end{eqnarray}
 where in the last line we have taken into account that $c G_N^{(D)} \ll r_h^{D-2}$. Thus with island configuration, at late time, entropy of Hawking radiation is bounded by black hole Bekenstein-Hawking entropy which decreases monotonically due to black hole evaporation.

\subsubsection{without island}
As in the case of $4D$, if there is no island, we can equivalently fix $r_I=0$ and $t_I$ as some constant. Then area term eq.(\ref{grav-d}) vanishes, we only have field term eq.(\ref{ent-d}). $u_{\text{in}}(I)=t_I$, $e^{2\rho(u_{\text{in}},v)(I)}=1$ since now $r_I=0$ is in the Minkowski region (\ref{Minkowski-d}), then
\begin{eqnarray}
S_{\text{gen}}&&=\frac{c}{12}  \ln \left( (u_{\text{in}}(A)-u_{\text{in}}(I))^2 (v_I-v_A)^2 e^{2\rho_A}
\right)_{t_{\pm}=0}\\
&&=\frac{c}{12}  \ln \left( (u_{\text{in}}(A)-t_I)^2 (v_I-v_A)^2 \frac{(1-\frac{r_h^{D-3}}{r_A^{D-3}})}{(D-3)^2 \exp((D-3)\frac{r^{*}_{>}(A)}{r_h}) } \frac{D-3}{e^{\frac{(D-3) v_0}{2 r_h}-\psi ^{(0)}\left(1+\frac{1}{D-3}\right)-\gamma }} \frac{(D-3)V_A}{2 r_h}
\right)\nb\\
\\
&&\approx\frac{c}{24r_h} (D-3) t_A,\lb{no-island-d}
\end{eqnarray}
where in the last line we take late time limit $t_A \gg r_A (\gg r_h)$.\footnote{There is also a logarithm divergent term proportional to $\ln(t_A)$ in eq.(\ref{no-island-d}), but it is only subdominant, so we omitted it.} Thus without island configuration, at late time radiation entropy grows linearly with time, which is consistent with Hawking's result. Compare eq.(\ref{no-island-d}) with eq.(\ref{island-d}), we have
  \begin{eqnarray}
S(R)=\text{min}( S_{\text{gen}}) \approx \text{min}(S_{\text{ent}},S_{\text{BH}}),
         \end{eqnarray}
 thus Page time is at
   \begin{eqnarray}\lb{Page-d}
 t_{\text{Page}}&\approx&\frac{24 r_h}{c (D-3)}S_{\text{BH}}.
         \end{eqnarray}
For higher dimensional black hole, our result for Page time is about twice as much as the Page time calculated in \cite{Hashimoto:2020cas} for Hartle-Hawking state of eternal black hole.

\subsection{Cutoff surface near horizon}

  \subsubsection{with island }
 We now consider the case in which cutoff surface is near horizon.\footnote{The calculation in this section  is similar to the one in \cite{Hashimoto:2020cas} apart from different convention of Kruskal coordinates and we do not assume $t_A=t_I$, but deduce it.} In this case we still have
 \bqn
S_{\text{gravity}}=\frac{\Omega_{D-2} r_I^{D-2}}{4G_N^{(D)}}.
\eqn
But entropy for fields is now given by
\begin{eqnarray}\lb{sentD}
S_{\text{ent}}&=&-\kappa_D c \frac{\text{Area}_D}{L^{D-2}},
\end{eqnarray}
where $L$ is geodesic distance between $\partial I$ and cutoff surface. Since we assume both $\partial I$ and cutoff surface are near horizon, $L$ can be approximately given by
\bqn
L&\approx&\sqrt{d(A,I)^2 e^{\rho_A} e^{\rho_I}}\nb\\
&=&\sqrt{\frac{\left(U_A-U_I\right) \left(-V_A+V_I\right)}{(D-3)^2} \sqrt{(1-\frac{r_h^{D-3}}{r_A^{D-3}})(1-\frac{r_h^{D-3}}{r_I^{D-3}})e^{-(D-3)\frac{r_>^*(A)+r_>^*(I)}{r_h}}}}\nb\\
\lb{eq4.69}
&\approx&\left(\frac{2 \sqrt{\beta } r_h}{\sqrt{D-3}}+O\left(\beta ^{3/2}\right)\right)+\sqrt{\alpha } \left(-\frac{e^{\frac{(-D-3) \left(t_I+t_A\right)}{2 r_h}} \left(e^{\frac{D t_I+3 t_A}{r_h}}+e^{\frac{3 t_I+D t_A}{r_h}}\right) r_h}{\sqrt{D-3}}+O\left(\beta ^1\right)\right)+O\left(\alpha ^1\right),\nb\\
 \eqn
where in the last line, we have made replacement $r_A=r_h(1+\alpha)$ and $r_I=r_h(1+\beta)$ expand $L$ to first order in $\alpha$ and zeroth order in $\beta$.\footnote{In \cite{Hashimoto:2020cas}, the authors further assume $t_A=t_I$ and eq.(\ref{eq4.69}) reduces to $\frac{2 \left(\sqrt{\beta }-\sqrt{\alpha }\right) r_h}{\sqrt{D-3}}$, which matches the result in \cite{Hashimoto:2020cas}. And this confirms that geodesic distance $L$ can be approximately given by $\sqrt{d(A,I)^2 e^{\rho_A} e^{\rho_I}}$ when $A$ and $\partial I$ is nearby. In this paper, we will not assume $t_A=t_I$, but we will deduce it in eq.(\ref{ta=ti}).} This is appropriate since we assume both $\partial I$ and $A$ both are near horizon. We will justify this assumption later. Then
\bqn
S_{\text{gen}}&&=\frac{\Omega_{D-2} r_I^{D-2}}{4G_N^{(D)}}-\kappa_D c \frac{\text{Area}_D}{L^{D-2}}\nb\\
\lb{sgen-out-d-near}
&&\approx \frac{\pi ^{\frac{D-1}{2}} \left(\frac{(\beta +1)^{D-2} r_h^D}{G_N^{(D)}}-c 2^{4-D} \kappa _D r_h^2 \left(\frac{\sqrt{\beta }-\sqrt{\alpha } \cosh \left(\frac{(D-3) \left(t_A-t_I\right)}{2 r_h}\right)}{(\alpha +1) \sqrt{D-3}}\right)^{2-D}\right)}{2 \Gamma \left(\frac{D-1}{2}\right) r_h^2}
\eqn
 Extremizing (\ref{sgen-out-d-near}) over $t_I$, we get
 \bqn
  \frac{\partial S_{\text{gen}}}{\partial t_I} \approx -\frac{\sqrt{\alpha } c 2^{2-D} (2-D) \sqrt{D-3} \pi ^{\frac{D-1}{2}} \kappa _D \sinh \left(\frac{(D-3) \left(t_A-t_I\right)}{2 r_h}\right) \left(\frac{\sqrt{\beta }-\sqrt{\alpha } \cosh \left(\frac{(D-3) \left(t_A-t_I\right)}{2 r_h}\right)}{(\alpha +1) \sqrt{D-3}}\right)^{1-D}}{(\alpha +1) \Gamma \left(\frac{D-1}{2}\right) r_h}=0,\nb\\
\eqn
which has solutions
\bqn\lb{ta=ti}
t_I=t_A, \; t_I=t_A-\frac{2 r_h \cosh ^{-1}\left(\sqrt{\frac{\beta }{\alpha }}\right)}{D-3},\; t_I=t_A+\frac{2 r_h \cosh ^{-1}\left(\sqrt{\frac{\beta }{\alpha }}\right)}{D-3}.
\eqn
Since $\beta<\alpha$, $\cosh ^{-1}\left(\sqrt{\frac{\beta }{\alpha }}\right)$ is imaginary. Dropping complex solution, we are left with
\bqn
t_I=t_A,
\eqn
which again confirms the assumption $t_I=t_A$ in \cite{Hashimoto:2020cas} in the higher dimensional case. Plug $t_I=t_A$ back into eq.(\ref{sgen-out-d-near}), we have
\bqn\lb{sgen-out-d-near-sim}
S_{\text{gen}} \approx \frac{\pi ^{\frac{D-1}{2}} \left(\frac{(\beta +1)^{D-2} r_h^D}{G_N^{(D)}}-c 2^{4-D} \kappa _D r_h^2 \left(\frac{\sqrt{\beta }-\sqrt{\alpha }}{(\alpha +1) \sqrt{D-3}}\right)^{2-D}\right)}{2 \Gamma \left(\frac{D-1}{2}\right) r_h^2}
\eqn
 Extremizing (\ref{sgen-out-d-near-sim}) over $\beta$, we get
 \bqn
  \frac{\partial S_{\text{gen}}}{\partial \beta} &&\approx \frac{\pi ^{\frac{D-1}{2}} \left(\frac{(D-2) (\beta +1)^{D-3} r_h^D}{G_N^{(D)}}-\frac{c 2^{3-D} (2-D) \kappa _D r_h^2 \left(\frac{\sqrt{\beta }-\sqrt{\alpha }}{(\alpha +1) \sqrt{D-3}}\right)^{1-D}}{(\alpha +1) \sqrt{\beta  (D-3)}}\right)}{2 \Gamma \left(\frac{D-1}{2}\right) r_h^2}\\
  \lb{eq4.76}
  &&\approx \frac{\pi ^{\frac{D-1}{2}} \left(\frac{(D-2)  r_h^D}{G_N^{(D)}}-\frac{c 2^{3-D} (2-D) \kappa _D r_h^2 \left(\frac{-\sqrt{\alpha }}{ \sqrt{D-3}}\right)^{1-D}}{ \sqrt{\beta  (D-3)}}\right)}{2 \Gamma \left(\frac{D-1}{2}\right) r_h^2}=0,
\eqn
where in the last line we have taken into account that $\beta<\alpha\ll1$. Eq.(\ref{eq4.76}) has solution\footnote{In the $D \rightarrow 4$ limit, eq.(\ref{beta-d}) will reduce to the $4D$ result (\ref{beta2}).}
\bqn\lb{beta-d}
\beta &&\approx c^2 4^{3-D} (D-3)^{D-2} (G_N^{(D)})^2 \alpha ^{1-D} \kappa _D^2 r_h^{4-2 D}\\
&&=2^{6-2D} (\kappa_D c G_N^{(D)})^2 (D-3)^{D-2} (a-r_h)^{1-D} r_h^{3-D},
\eqn
which matches the result in \cite{Hashimoto:2020cas}, and confirms the assumption that $\partial I$ is near and outside horizon due to the fact that $c G_N^{(D)} \ll r_h^{D-2}$.   Plug eq.(\ref{beta-d}) into (\ref{sgen-out-d-near-sim}) and expand the result to zeroth order in $G_N^{(D)}$, we have
     \bqn
S_{\text{gen}}&&= \frac{\pi ^{\frac{D-1}{2}} r_h^{D-2}}{2 G_N^{(D)} \Gamma \left(\frac{D-1}{2}\right)}-\frac{ (-1)^D 2^{3-D}c \kappa _D \pi ^{\frac{D-1}{2}} \left(\frac{\alpha }{D-3}\right)^{1-\frac{D}{2}} (\alpha +1)^{D-2} }{\Gamma \left(\frac{D-1}{2}\right)}+O\left(G^1\right)\\
&&\approx \frac{\pi ^{\frac{D-1}{2}} r_h^{D-2}}{2 G_N^{(D)} \Gamma \left(\frac{D-1}{2}\right)}=S_{\text{BH}},\lb{island2-d}
\eqn
where in the last line we have taken into account that $c G_N^{(D)} \ll r_h^{D-2}$. Thus for the case that cutoff surface near horizon, with island configuration, at late time, entropy of Hawking radiation is bounded by black hole Bekenstein-Hawking entropy which decreases monotonically due to black hole evaporation.

    \subsubsection{without island}
  On the other hand, in the case without island, we still need to use eq.(\ref{ent-d}) to calculate $S_{\text{ent}}$ due to the fact that $r_I=0$ is far from near-horizon cutoff surface for macroscopic black hole. Then the result of $S_{\text{gen}}$ without island for near-horizon cutoff surface will be same as eq.(\ref{no-island-d}). Comparing with (\ref{island2-d}), we find same Page time
     \begin{eqnarray}\lb{Page2-d}
 t_{\text{Page}}&\approx&\frac{24 r_h}{c (D-3)}S_{\text{BH}}.
         \end{eqnarray}

\section{Conclusion and discussion}\lb{discussion}
 \renewcommand{\theequation}{6.\arabic{equation}}\setcounter{equation}{0}
In this paper, unlike other papers that discuss island rule of eternal black hole, we make a further step towards more realistic black hole that formed from gravitational collapse. Based on this motivation, we choose ``in'' vacuum state $|in\rangle$ which describes vacuum of Minkowski region and has thermal Hawking flux in Schwarzschild region.  Due to the absence of incoming flux, $s$-wave approximation is valid for ``in'' vacuum state of one-sided black hole formed from collapsing of spherical null shell when cutoff surface $A$ is far from horizon. The use of ``in'' vacuum state is the key difference between our paper and the other papers that use Hartle-Hawking state \cite{Alishahiha:2020qza,Hashimoto:2020cas,Matsuo:2020ypv,Arefeva:2021kfx,Kim:2021gzd,Wang:2021woy,He:2021mst,Matsuo:2021mmi}. 

We find that, when cutoff surface is far from horizon, island emerges at late time with its boundary $\partial I$ inside and at vicinity of horizon and saves the entropy bound. Page time is about twice as much as the one of Hartle-Hawking state. On the other hand, when cutoff surface is near horizon, island emerges at late time with its boundary $\partial I$ outside and at vicinity of horizon and saves the entropy bound. Page time is also about twice as much as the one of Hartle-Hawking state. For $D>4$ dimensional case, we find similar results.

Our result is different from the case of eternal black hole in which boundary of island $\partial I$ always emerge outside horizon no matter cutoff surface is far from or near horizon. Different states manifestly affect $S_{\text{ent}}$ in island formula when cutoff surface is far from horizon and thus have different result for Page time and different position of boundary of island $\partial I$.

\section*{Acknowledgments}
 W-C.G. is supported by Baylor University through the Baylor Physics graduate program.   This work is partially supported by the National Natural Science
Foundation of China with Grant No. 11975116, and the Jiangxi Science Foundation for
Distinguished Young Scientists under Grant No. 20192BCB23007.

  \begin{appendix}
  \section{Hawking radiation in the background of dynamical black hole and $s$-wave approximation}\lb{app-1}
 \renewcommand{\theequation}{A.\arabic{equation}}\setcounter{equation}{0}
For simplicity, we consider massless scalar fields in $4D$ spacetimes. The fields satisfy Klein-Gordon (KG) equation
 \bqn
 \square f=0.
 \eqn
  In spherical symmetric spacetimes, we can expand the fields in terms of spherical harmonic functions
 \bqn
 f(x^{\mu})=\sum_{\ell,m}\frac{f_{\ell}(t,r)}{r}Y_{\ell m}(\theta,\phi).
 \eqn
 Then in Schwarzschild spacetime, KG equation reduces to two-dimensional equation for $f_{\ell}(t,r)$
  \bqn
 \left( -\frac{\partial^2}{\partial t^2}+\frac{\partial^2}{\partial {r^*}^2}-V_{\ell}(r)\right) f_{\ell}(t,r)=0,
 \eqn
 where the potential barrier is given by
 \bqn
 V_{\ell}(r)=\left(1-\frac{r_h}{r}\right) \left(\frac{\ell (\ell+1)}{r^2}+\frac{r_h}{r^3}\right).
 \eqn
 The potential barrier vanishes at $r=+\infty$ ($r^*=+\infty$) and at the event horizon $r=r_h$ ($r^*=-\infty$). So fields can be treated as effective two-dimensional free massless fields at $r=+\infty$ and  $r=r_h$. And $V_{\ell}(r)$ grows like $\ell^2$, so the outgoing Hawking radiation which propagates from horizon $r=r_h$ to distant observer at $r=+\infty$ (or $r \gg r_h$) is dominated by the modes with lowest angular momentum, i.e. $s$-wave ($\ell=0$) component \cite{Fabbri:2005mw,Harlow:2014yka} since higher angular momentum modes are more likely back-scattered by the potential barrier. 
 
 This is different from the case in Minkowski spacetime where KG equation reduces to
 \bqn
 \left( -\frac{\partial^2}{\partial t^2}+\frac{\partial^2}{\partial r^2}-\frac{\ell (\ell+1)}{r^2}\right) f_{\ell}(t,r)=0,
 \eqn
where the potential $\frac{\ell (\ell+1)}{r^2}$ does not vanish as $r \rightarrow 0$. So $s$-wave approximation is not valid in Minkowski spacetime. This also can be seen from the fact that in Minkowski spacetime, in odd spacetime dimensions, region with spherical entangling surface has entanglement entropy without logarithmic term \cite{Ryu:2006ef}. If $s$-wave approximation is valid, then entanglement entropy should have logarithmic term like Eq.\eqref{sent}.
 
 Moreover, for eternal black hole in Hartle-Hawking state, in addition to outgoing modes, there are also incoming modes propagating from past null infinity $\mathcal{J}^-$ in Schwarzschild region and part of the modes with high angular momentum will be back-scattered to future null infinity $\mathcal{J}^+$ \cite{Wald:1984rg}. Thus in this case, distant observer will also receive modes with high angular momentum, so the $s$-wave approximation is not valid. This also can be seen from the fact that in Hartle-Hawking state in  Schwarzschild spacetime, in odd spacetime dimensions, region with spherical entangling surface has entanglement entropy without logarithmic term \cite{Solodukhin:2011gn}.
 
 We would like to emphasize that in this paper, we only consider ``in'' vacuum state in one-sided black hole formed by spherical null shell collapsing at $v=v_0$. There are only outgoing modes in such state  and no incoming flux from $\mathcal{J}^-$ in Schwarzschild region \cite{Fabbri:2005mw,Harlow:2014yka}, and this is true for both odd and even spacetime dimensions. So we would assume in both odd and even spacetime dimensions, $s$-wave approximation is valid for distant observer (cutoff surface far from horizon) in our paper. Then the fields can be treated as effective $(1+1)$ dimensional massless fields in $(t,r)$ direction \cite{Fabbri:2005mw,Harlow:2014yka}. The use of ``in'' vacuum state is the key difference between our paper and the other papers that use Hartle-Hawking state \cite{Alishahiha:2020qza,Hashimoto:2020cas,Matsuo:2020ypv,Arefeva:2021kfx,Kim:2021gzd,Wang:2021woy,He:2021mst,Matsuo:2021mmi}. We would like to leave rigorous discussion on the validity of $s$-wave approximation for future work.

\section{Cutoff surface near horizon with boundary of island inside horizon}\lb{app-a}
 \renewcommand{\theequation}{B.\arabic{equation}}\setcounter{equation}{0}

 For the case that $\partial I$ is inside horizon, we need to use (\ref{Kruskal-in})
 and thus $L$ can be approximately given by
\bqn
L&=&\sqrt{d(A,I)^2 e^{\rho_A} e^{\rho_I}}\nb\\
&=&\sqrt{\left(U_A-U_I\right) \left(-V_A+V_I\right) \sqrt{\frac{r_h^2 e^{-\frac{r_A}{r_h}-\frac{r_I}{r_h}}}{r_I r_A}}}\nb\\
&=&\sqrt{\frac{4 r_h^3 e^{-\frac{r_A+t_A+r_I+t_I}{2 r_h}} \left(\sqrt{\frac{r_A}{r_h}-1} e^{\frac{r_A+t_I}{2 r_h}}+\sqrt{1-\frac{r_I}{r_h}} e^{\frac{t_A+r_I}{2 r_h}}\right) \left(\sqrt{\frac{r_A}{r_h}-1} e^{\frac{r_A+t_A}{2 r_h}}-\sqrt{1-\frac{r_I}{r_h}} e^{\frac{r_I+t_I}{2 r_h}}\right)}{\sqrt{r_I r_A}}}\nb\\
\eqn

  Then we have
\bqn
S_{\text{gen}}&=&\frac{\pi r_I^2}{G_N}-\kappa c \frac{\text{Area}}{L^2}\\
&=&\frac{\pi  c \kappa  \sqrt{r_I r_A} r_A^2 e^{\frac{r_A+t_A+r_I+t_I}{2 r_h}}}{r_h^3 \left(\sqrt{\frac{r_A}{r_h}-1} e^{\frac{r_A+t_I}{2 r_h}}+\sqrt{1-\frac{r_I}{r_h}} e^{\frac{t_A+r_I}{2 r_h}}\right) \left(-\sqrt{\frac{r_A}{r_h}-1} e^{\frac{r_A+t_A}{2 r_h}}+\sqrt{1-\frac{r_I}{r_h}} e^{\frac{r_I+t_I}{2 r_h}}\right)}+\frac{\pi  r_I^2}{G_N}\nb\\
\eqn

  Since we assume both $\partial I$ and cutoff surface are near horizon, we can make replacement $r_A=r_h(1+\alpha)$ and $r_I=r_h(1-\beta)$, where $\alpha,\beta \ll 1$,
 then we have
 \bqn\lb{sgen-in2}
S_{\text{gen}}= \frac{\pi  (\beta -1)^2 r_h^2}{G_N}-\frac{\pi  (\alpha +1)^{5/2} \sqrt{1-\beta } c \kappa }{(\alpha +\beta ) \sinh \left(\frac{\alpha +\beta }{2}\right)+(\alpha -\beta ) \cosh \left(\frac{\alpha +\beta }{2}\right)+2 \sqrt{\alpha  \beta } \sinh \left(\frac{t_A-t_I}{2 r_h}\right)}.\nb\\
  \eqn
   Extremizing (\ref{sgen-in2}) over $t_I$, we get
 \bqn
  \frac{\partial S_{\text{gen}}}{\partial t_I}=-\frac{\pi  (\alpha +1)^{5/2} \sqrt{1-\beta } c \kappa  \sqrt{\alpha  \beta } \cosh \left(\frac{t_A-t_I}{2 r_h}\right)}{r_h \left((\alpha +\beta ) \sinh \left(\frac{\alpha +\beta }{2}\right)+(\alpha -\beta ) \cosh \left(\frac{\alpha +\beta }{2}\right)+2 \sqrt{\alpha  \beta } \sinh \left(\frac{t_A-t_I}{2 r_h}\right)\right)^2}=0,
  \eqn
  which has solutions
  \bqn
 t_I=t_A-i \pi r_h (4 c_1-1),\; c_1\in \mathbb{Z},
  \eqn
  or
  \bqn
  t_I=t_A-i \pi r_h (4 c_1+1),\; c_1\in \mathbb{Z},
  \eqn
where $\mathbb{Z}$ represents integers. Since these are complex solutions, they have no physical meaning and need to be dropped.

In higher dimensional case, we have similar result that there is no physical solution of $\partial I$ inside horizon when cutoff surface is near horizon.

\section{Higher dimensional connection condition near horizon}\lb{apx.b}
 \renewcommand{\theequation}{B.\arabic{equation}}\setcounter{equation}{0}

  \subsection{Outside and near horizon}

 To have smooth metric at the null shell, we have connecting condition outside horizon
    \begin{eqnarray}
  u &=& \frac{2 r^{D-2} r_h^{3-D} \, _2F_1\left(1,1+\frac{1}{D-3};2+\frac{1}{D-3};\left(\frac{r}{r_h}\right){}^{D-3}\right)}{D-2}+\frac{2 i \pi  r_h}{D-3}+v_0\\
  \lb{u-D-out-near}
  & \approx&\frac{2 r_h^{3-D} \, _2F_1\left(1,1+\frac{1}{D-3};2+\frac{1}{D-3};(\epsilon +1)^{D-3} \right) \left((\epsilon +1) r_h\right){}^{D-2}}{D-2}+\frac{2 i \pi  r_h}{D-3}+v_0\\
 & \approx&  -\frac{2 \ln (3-D) r_h+2 \psi ^{(0)}\left(1+\frac{1}{D-3}\right) r_h-D v_0+2 r_h \ln (\epsilon )+2 i \pi  r_h+2 \gamma  r_h+3 v_0}{D-3}.
    \end{eqnarray}
    where we have assumed $r/r_h=\frac{v_0- u_{\text{in}}}{2 r_h}=1+\epsilon$, and expanded eq.(\ref{u-D-out-near}) around $\epsilon=0$. Solve $u_{\text{in}}$ in terms of $u$. Then
        \begin{eqnarray}\lb{uin-d}
    u_{\text{in}} \approx r_h \left(-\frac{2 \exp \left(-\frac{(D-3) \left(u-v_0\right)}{2 r_h}-\psi ^{(0)}\left(1+\frac{1}{D-3}\right)-\gamma \right)}{D-3}-2\right)+v_0,
       \end{eqnarray}
eq.(\ref{uin-d}) is only valid outside horizon.

Since outside horizon we have
  \begin{eqnarray}
   u=\frac{2 r_h}{-(D-3)} \ln \frac{U}{-2 r_h},
   \end{eqnarray}
   then
     \begin{eqnarray} \lb{uin-U-d}
 u_{\text{in}} \approx  \frac{U e^{\frac{(D-3) v_0}{2 r_h}-\psi ^{(0)}\left(1+\frac{1}{D-3}\right)-\gamma }}{D-3}-2 r_h+v_0,
      \end{eqnarray}
   and thus
       \begin{eqnarray}\lb{derivative-d2}
    \frac{dU}{d u_{\text{in}}} \simeq \frac{D-3}{e^{\frac{(D-3) v_0}{2 r_h}-\psi ^{(0)}\left(1+\frac{1}{D-3}\right)-\gamma }}, \; \frac{dV}{d v}=\frac{(D-3)V}{2 r_h},
        \end{eqnarray}
             thus
        \begin{eqnarray}
e^{2\rho(u_{\text{in}},v)}&=&e^{2\rho(U,V)}\frac{dU}{d u_{\text{in}}}\frac{dV}{d v}\\
&\approx& e^{2\rho(U,V)} \frac{D-3}{e^{\frac{(D-3) v_0}{2 r_h}-\psi ^{(0)}\left(1+\frac{1}{D-3}\right)-\gamma }} \frac{(D-3)V}{2 r_h}\\
&=&\frac{(1-\frac{r_h^{D-3}}{r^{D-3}})}{(D-3)^2 \exp((D-3)\frac{r^{*}_{>}}{r_h}) } \frac{D-3}{e^{\frac{(D-3) v_0}{2 r_h}-\psi ^{(0)}\left(1+\frac{1}{D-3}\right)-\gamma }} \frac{(D-3)V}{2 r_h},
\end{eqnarray}

 \subsection{Inside and near horizon}
 The connecting condition inside horizon is\footnote{Here we still define $u=t-r^*$ inside horizon.}
     \begin{eqnarray}
  u &=& \frac{2 r^{D-2} r_h^{3-D} \, _2F_1\left(1,1+\frac{1}{D-3};2+\frac{1}{D-3};\left(\frac{r}{r_h}\right){}^{D-3}\right)}{D-2}+v_0,\\
  \lb{u-D-in-near}
  & \approx&\frac{2 r_h^{3-D} \, _2F_1\left(1,1+\frac{1}{D-3};2+\frac{1}{D-3};(1-\epsilon)^{D-3}  \right) \left((1-\epsilon ) r_h\right){}^{D-2}}{D-2}+v_0\\
   & \approx&  -\frac{2 \ln (D-3) r_h+2 \psi ^{(0)}\left(1+\frac{1}{D-3}\right) r_h-D v_0+2 r_h \ln (\epsilon )+2 \gamma  r_h+3 v_0}{D-3}
    \end{eqnarray}
   where we have assumed $r/r_h=\frac{v_0- u_{\text{in}}}{2 r_h}=1-\epsilon$, and expanded eq.(\ref{u-D-in-near}) around $\epsilon=0$. Solve $u_{\text{in}}$ in terms of $u$. Then
        \begin{eqnarray}\lb{uin-d-in}
    u_{\text{in}} \approx r_h \left(\frac{2 \exp \left(-\frac{(D-3) \left(u-v_0\right)}{2 r_h}-\psi ^{(0)}\left(1+\frac{1}{D-3}\right)-\gamma \right)}{D-3}-2\right)+v_0,
       \end{eqnarray}
eq.(\ref{uin-d-in}) is only valid inside horizon.

   Since inside horizon we have
  \begin{eqnarray}
   u=\frac{2 r_h}{-(D-3)} \ln \frac{U}{2 r_h},
   \end{eqnarray}
  then
    \begin{eqnarray} \lb{uin-U-in-d}
u_{\text{in}} \approx  \frac{U e^{\frac{(D-3) v_0}{2 r_h}-\psi ^{(0)}\left(1+\frac{1}{D-3}\right)-\gamma }}{D-3}-2 r_h+v_0,
     \end{eqnarray}
     which is the same as eq.(\ref{uin-U-d}).

     Then
       \begin{eqnarray}\lb{derivative-d3}
    \frac{dU}{d u_{\text{in}}} \simeq  \frac{D-3}{e^{\frac{(D-3) v_0}{2 r_h}-\psi ^{(0)}\left(1+\frac{1}{D-3}\right)-\gamma }}, \; \frac{dV}{d v}=\frac{(D-3)V}{2 r_h},
        \end{eqnarray}

          thus
        \begin{eqnarray}
e^{2\rho(u_{\text{in}},v)}&=&e^{2\rho(U,V)}\frac{dU}{d u_{\text{in}}}\frac{dV}{d v}\\
&\approx& e^{2\rho(U,V)} \frac{D-3}{e^{\frac{(D-3) v_0}{2 r_h}-\psi ^{(0)}\left(1+\frac{1}{D-3}\right)-\gamma }} \frac{(D-3)V}{2 r_h}\\
&=&\frac{(1-\frac{r_h^{D-3}}{r^{D-3}})}{(D-3)^2 \exp((D-3)\frac{r^{*}_{>}}{r_h}) } \frac{D-3}{e^{\frac{(D-3) v_0}{2 r_h}-\psi ^{(0)}\left(1+\frac{1}{D-3}\right)-\gamma }} \frac{(D-3)V}{2 r_h},\\
\lb{in-near-d}
&=&-\frac{(1-\frac{r_h^{D-3}}{r^{D-3}})}{(D-3)^2 \exp((D-3)\frac{r^{*}_{<}}{r_h}) } \frac{D-3}{e^{\frac{(D-3) v_0}{2 r_h}-\psi ^{(0)}\left(1+\frac{1}{D-3}\right)-\gamma }} \frac{(D-3)V}{2 r_h}.
\end{eqnarray}
  \end{appendix}

\end{document}